\documentclass{iopart}
\usepackage{iopams,epsfig,graphicx} 
\newcommand{\lapx}{\,\raisebox{-.5ex}
 {$\stackrel{\raisebox{-.5pt}{$\textstyle <$}}{\sim}$}\,}
\newcommand{\gapx}{\,\raisebox{-.5ex}
 {$\stackrel{\raisebox{-1pt}{$\textstyle >$}}{\sim}$}\,}
 
 \newcommand{\entspricht}{\stackrel{\scriptscriptstyle\,\wedge}{=}}

\begin{document}

\title[Dynamics of dissipative coupled spins: decoherence and relaxation]{Dynamics of dissipative coupled spins:
decoherence, relaxation and effects of a spin-boson bath}

\author{P. N\"agele, G. Campagnano, and U. Weiss\footnote{Author to whom any correspondence should be addressed.}}

\address{II. Institut f\"ur Theoretische Physik, Universit\"at Stuttgart,
 D-70550 Stuttgart, Germany.}
 \ead{naegele@theo2.physik.uni-stuttgart.de,
      campagnano@theo2.physik.uni-stuttgart.de, and
          weiss@theo2.physik.uni-stuttgart.de}
\begin{abstract}
We study the reduced dynamics of interacting spins, each coupled to its own bath of bosons. We derive the solution 
in analytic form in the white-noise limit and analyze the rich behaviors in diverse limits ranging from weak coupling 
and/or low temperature to strong coupling and/or high temperature. We also view the one spin as being coupled to a 
spin-boson environment and consider the regimes in which it is effectively nonlinear, and in which it can be regarded 
as a resonant bosonic environment.

\end{abstract}

\section{Introduction}

Comprehension of the phenomenon of {\em decoherence} in open quantum systems has always attracted 
much attention, in particular as a prerequisite to understand the transition from 
quantum to classical behavior. The dissipative two-state or {\em spin-boson} model has been thoroughly studied in wide regions
of the parameter space with diverse methods and techniques since the 80's \cite{leggett,bookweiss}.

In the last decade, the subject of decoherence has experienced renaissance following the growing  interest in the field of quantum
state manipulation and quantum computation \cite{nielsen-chuang}. 
Any noise source sensitively leads to a narrowing of the quantum coherence domain. This entails severe limitations
for coupled qubits  to perform logic quantum operations. For this reason, extensive understanding of the decoherence mechanisms
is indispensable.

In this work, we focus upon a model which is a generalization of the single spin-boson model to the  case of two spins  
which mutually interact via an Ising-type coupling and are coupled to independent environments made up by bosons.
The first analysis of this model relying on the influence functional method was given by 
Dub\'{e} and Stamp \cite{stamp}. They obtained results for the dynamics in analytic form in restricted regions of the 
parameter space by omitting certain classes of path contributions and bath correlations.
Several other previous studies on the same or related models relied on the master equation and/or perturbative 
Redfield approach \cite{governale,thorwart,wilhelm}.
Besides the weak-coupling assumption, often the secular approximation \cite{blum} is made, which breaks down 
however when the spectrum becomes degenerate. 

The model allows, for instance, to study decoherence and relaxation of two coupled qubits \cite{governale,wilhelm}, 
or the influence of a bistable impurity on the qubit dynamics \cite{paladino08}.  
The latter may  significantly degrade coherence in Josephson phase qubits \cite{simmonds}. 
Other possible application is study of coherence effects in coupled molecular magnets \cite{troiani}. 

In earlier works, the model has been analyzed in the pure dephasing regime both by the Feynman-Vernon method \cite{paladino04}
and the Lindblad approach \cite{paladino06}. Here we extend the work in Ref. \cite{paladino04} beyond the pure dephasing 
regime and include the full dynamics of the qubit. In particular, we are interested in the competition between decoherence and relaxation to the equilibrium state. Here we focus on the white-noise regime. We shall derive the exact solution for the 
reduced density matrix without restriction on the parameters of the model and analyze it in the coherent and incoherent domains
and in the crossover regions in between. 

The model and relevant quantities of the reduced dynamics are introduced in section 2. Section 3 deals with the exact 
formal solution for the reduced dy\-na\-mics. In section 4, the path sum is carried out in the white-noise domain
without any further approximation,  and analytic expressions for the relevant expectation values in Laplace space
are presented. After an overview of the qualitative features of the dynamics in section 5, we  present in section 6
explicit expressions for decoherence and relaxation in the various parameter regimes ranging from low temperture and/or
weak coupling to high temperature and/or strong coupling. Finally, we study in section 7 the influence of a nonlinear spin-boson environment on the second spin in the various limits. We demonstrate that it behaves in the weak-coupling limit 
as a bosonic (linear) bath with a resonant spectral structure. 

\section{Model}
We consider two two-state systems which are coupled to each other via an Ising-type coupling and to independent 
bosonic environments. In pseudospin representation, we choose the generalized spin-boson Hamiltonian 
(we use units where $\hbar =k_{\rm B}^{} =1$)
\begin{equation}\label{ham1}
\fl\quad
H\;=\;-\frac{\Delta_1}{2}\sigma_x \,-\, \frac{\Delta_2}{2} \tau_x
\,-\, \frac{v}{2}\sigma_z \tau_z\,-\,\frac{1}{2}\sigma_z X_1 \,-\,\frac{1}{2}\tau_z X_2  \,+\, 
\sum_{\zeta=1,2}\sum_\alpha \omega_{\zeta,\alpha}^{} b_{\zeta,\alpha}^\dag b_{\zeta,\alpha}^{}    \;.
\end{equation}
In the basis formed by the localized eigen states $|R\!\!>$ and $|L\!\! >$ of $\sigma_z$ and $\tau_z$, respec\-tively, 
$\Delta_1$ and $\Delta_2$ represent the tunneling couplings between the localized states,
and the coupling term $ -\, \frac{1}{2}v \sigma_z \tau_z$ acts as a mutual 
bias energy of strength $v$. The collective bath modes  
$X_\zeta(t) = \sum_{\alpha} c_{\zeta,\alpha}^{} [\,b_{\zeta,\alpha}^{}(t)\,+\,b_{\zeta,\alpha}^\dag(t)\,]$ 
($\zeta = 1,2$) represent fluctuating bias forces.
The Hamiltonian is very rich in content and may model diverse physical situations. It may describe two coupled qubits or a qubit $\sigma$
in contact with a complex environment formed by a bistable dissipative impurity $\tau$. Other possible realizations are coupled molecular
magnets of which the low-energy states can be viewed as a spin \cite{troiani}.

For the model (\ref{ham1}), all effects of the environments are captured by the power spectrum of the collective
bath modes
\begin{equation}\label{powerspec}
\fl\;\;
S_{\zeta,\zeta'}^{}(\omega) \;=\; \frac{1}{2}\int_{-\infty}^\infty\!\!\!{\rm d}t\,
\left\langle X_{\zeta}^{}(t)X_{\zeta'}^{}(0)\,+\,X_{\zeta'}^{}(0)X_{\zeta}^{}(t)\right\rangle_\beta \;=\; 
\delta_{\zeta,\zeta'}^{}\,\pi G_\zeta^{}(\omega)\coth\Big(\frac{\beta\omega}{2}\Big)
\end{equation}
with the spectral density of the coupling \cite{leggett,bookweiss}
\begin{equation}\label{specdens}
\fl\qquad
G_\zeta(\omega)\;=\; \sum_\alpha  c^2_{\zeta\!,\,\alpha} \delta(\omega - \omega_{\zeta\!,\,\alpha} )
\;=\; 2 K_\zeta\, \omega \,{\rm e}^{-|\omega| /\omega_c}_{}  \;,\qquad \zeta =1,\,2 \;.
\end{equation}
Here, the second form represents the Ohmic case with a high-frequency cut-off $\omega_{\rm c}^{}$.
Alternatively, one may choose that the two spins are coupled to a common bath \cite{wilhelm}. Here we study the 
effects of independent environments. This case is realistic in most physical systems of actual interest.

The density matrix of a single spin has four matrix elements, the two populations that we shall label as ${\rm RR} \equiv 1$ and 
${\rm LL} \equiv 3$, and the two coherences with labels  ${\rm LR} \equiv 2$ and ${\rm RL} \equiv 4$. 
The two-spin density matrix has 16 matrix elements $\rho_{n,m}^{}(t)$. We choose for convenience that 
the first (second) index refers to the states
$n= 1,\cdots,4$ ($m=1,\cdots,4$) of the $\sigma$--spin ($\tau$--spin).
The matrix elements can be expressed in terms of expectation values of 15 operators,
$\langle \sigma_i\otimes \mathbf{1} \rangle_t = \langle \sigma_i\rangle_t$, 
$\langle \mathbf{1}\otimes\tau_i\rangle_t = \langle \tau_i\rangle_t $ and 
$\langle \sigma_i\otimes\tau_j\rangle_t = \langle \sigma_i\tau_j\rangle_t$
($i=1,\,2,\,3$ and $j=1,\,2,\,3$). The 4 pure populations may then be written as
\begin{equation}
\begin{array}{rcl}
\rho_{\rm 1,1}^{}(t) &=& [\, 1 \,+\,\langle\sigma_z^{}\rangle_t \,+\,\langle\tau_z^{}\rangle_t
\,+\,\langle\sigma_z^{}\tau_z^{}\rangle_t\,]/4 \;, \\[3mm]
\rho_{\rm 1,3}^{}(t) &=& [\, 1 \,+\,\langle\sigma_z^{}\rangle_t \,-\,\langle\tau_z^{}\rangle_t
\,-\,\langle\sigma_z^{}\tau_z^{}\rangle_t\,]/4 \;, \\[3mm]
\rho_{\rm 3,1}^{}(t) &=& [\, 1 \,-\,\langle\sigma_z^{}\rangle_t \,+\,\langle\tau_z^{}\rangle_t
\,-\,\langle\sigma_z^{}\tau_z^{}\rangle_t \,]/4\;, \\[3mm]
\rho_{\rm 3,3}^{}(t) &=& [\, 1 \,-\,\langle\sigma_z^{}\rangle_t \,-\,\langle\tau_z^{}\rangle_t
\,+\,\langle\sigma_z^{}\tau_z^{}\rangle_t \,]/4\;. 
\end{array}
\end{equation}
Corresponding expressions hold for the 4 pure coherences and the 8 hybrid states. For instance, we have
\begin{equation}
\rho_{\rm 2,4}^{}(t) \;=\; [\, \langle\sigma_x^{}\tau_x^{}\rangle_t \,+\,\langle\sigma_y^{}\tau_y^{}\rangle_t 
\,+\,{\rm i}\,\langle\sigma_x^{}\tau_y^{}\rangle_t \,-\,{\rm i}\,\langle\sigma_y^{}\tau_x^{}\rangle_t \,]/4 \; . 
\end{equation}
Here we are predominantly interested in the populations. Throughout we will choose that the reduced system starts out from
the initial state $\rho_{1,1}^{}(t=0) = 1$ while the heat reservoirs are in thermal equilibrium at temperature $T$.

In the absence of the environment, the Hamiltonian $H= H_0$ can be easily transformed into diagonal form
\begin{equation} \label{hamdiag}
\widetilde H_0 \;=\; -\frac{\Omega}{2} \,(\sigma_z \otimes \mathbf{1}) - \frac{\delta}{2} \, (\mathbf{1} \otimes \tau_z) \;.
\end{equation}
The eigenfrequencies are
\begin{equation} \label{def_eigen1}
 \Omega\; =\; \frac{1}{2} ( \,\Omega_{+} \, + \, \Omega_{-}\, ) \; , \quad \quad  
\delta \;=\; \frac{1}{2} (\, \Omega_{+} \,-\, \Omega_{-} \,) \; ,
\end{equation}
\begin{equation}\label{def_eigen2}
\fl\mbox{with}\qquad \qquad\Omega_\pm \;=\; \sqrt{(\Delta_1 \pm \Delta_2)^2+v^2}  \, ,
\end{equation}
and they obey the Vieta relations
\begin{equation}\label{vieta}
\fl
\begin{array}{rcl}
\Omega^2_{} +\delta^2_{} &=& \Delta_1^2+\Delta_2^2+v^2_{}\; ,\qquad \;\;\;
\Omega^2_{}\delta^2_{}\;=\;\Delta_1^2\Delta_2^2 \;, \\[3mm]
\Omega_-^2+\Omega_+^2 &=& 2(  \Delta_1^2+\Delta_2^2+v^2_{} )\;, \quad\;
\Omega_+^2\Omega_-^2 \;=\; (\Delta_1^2+\Delta_2^2+v^2_{})^2_{}- 4\Delta_1^2\Delta_2^2 \; . 
\end{array}
\end{equation}

The Liouville equations $\dot W_j(t) =-i [\,H,\,W_j(t)\,]$ ($j=1,\cdots,15$),  where the set $\{W_j(t)\}$ represents 
the above 15 operators,
yield 15 coupled equations. These are conveniently solved in Laplace space. For instance, we get
\begin{equation} \label{expec1}
\fl\;\;
 \begin{array}{rcl}
\langle \sigma_z (\lambda) \rangle &=& {\displaystyle \frac{\lambda (v^2 + \Delta_2^2 + \lambda^2)}{
(\lambda^2+\Omega^{2})(\lambda^2+\delta^{2})}   }\;, \qquad\quad
\langle \tau_z (\lambda) \rangle \;=\; {\displaystyle \frac{\lambda (v^2 + \Delta_1^2 + \lambda^2)}{
(\lambda^2+\Omega^{2})(\lambda^2+\delta^{2})}   }  \;,    \\[4mm]
  \langle \sigma_z\tau_z (\lambda) \rangle &=& {\displaystyle \frac{(v^2 + \lambda^2) (v^2 + \Delta_1^2 + \Delta_2^2 + \lambda^2)}{\lambda\, (\lambda^2+\Omega_{+}^{2})(\lambda^2+\Omega_{-}^{2})} }\; .
 \end{array}
\end{equation}
 
Here we are interested in the evolution of the two-spin system without restricting ourselves to weak damping. 
Therefore, we refrain from employing the perturbative Redfield approach.
Rather we calculate the reduced dynamics  with use of the Feynman-Vernon influence functional method.
We show that the solution is available in analytic form in 
the white noise limit for general parameters $\Delta_1^{},\, \Delta_2^{},\, v$ and $T$. 

\section{Formal solution for the reduced density matrix}
Within the Feynman-Vernon method, the exact formal expression for the RDM of the two-spin system is the quadruple 
path integral
\begin{equation}\label{rdm}
\fl
 \rho_{\sigma_{\rm f}^{} \sigma'_{\rm f},\tau_{\rm f}^{} \tau'_{\rm f}}^{}(t) = 
\int \mathcal{D}\sigma\,\mathcal{D}\sigma'\, \mathcal{D}\tau\, \mathcal{D}\tau'\; 
 {\mathcal A}[\sigma]\, {\mathcal A}^* [\sigma']\,
{\mathcal A}[\tau]\, {\mathcal A}^* [\tau'] \,{\cal B}[\sigma,\sigma';\tau,\tau']
\,\mathcal{F}[\sigma,\sigma';\tau,\tau']
\end{equation}
with appropriately chosen boundary values for the spin paths. Here, each of the
paths $\sigma(t'),\sigma'(t'),\tau(t') ,\tau'(t')$ starts out from the localized state $|R\!\!>$
at time zero. They end up at time $t$ in the states $|\sigma_{\rm f}^{}\!\!>,\,
|\sigma'_{\rm f}\!\!>,\,| \tau_{\rm f}^{}\!\!>$, and $| \tau'_{\rm f}\!\!>$, respectively, where
$\sigma_{\rm f}^{},\, \sigma'_{\rm f},\,\tau_{\rm f}^{},\, \tau'_{\rm f}\,\in\, {\rm R,\,L}$.
The functional ${\mathcal A}[\sigma]$ is the amplitude for the free spin 
$\sigma$ to follow the path $\sigma(t')$, the functional ${\cal B}[\sigma,\sigma';\tau,\tau']$ represents the 
coupling of the two spins (see below), and the functional $\mathcal{F}[\sigma,\sigma';\tau,\tau']$ 
introduces the environmental influences.

For uncorrelated baths, we have 
$ \mathcal{F}[\sigma,\sigma';\tau,\tau']\,=\,\mathcal{F}_1[\xi_1^{},\eta_1^{}]\,\,\mathcal{F}_2[\xi_2^{},\eta_2^{}]$,
where
\begin{equation} 
\fl \ln \mathcal{F}_\zeta [\xi_\zeta ,\eta_\zeta] = \int_0^t \!\!\!dt' \int_0^{t'}\!\!\! dt''
\left[ 
\dot{\xi}_\zeta (t')Q_\zeta'(t'-t'')\dot{\xi}_\zeta(t'') 
\,+\,{\rm i}\,\dot{\xi}_\zeta (t')Q_\zeta''(t'-t'')\dot{\eta}_\zeta(t'')
\right] \;. 
\end{equation} 
Here we have introduced symmetric and antisymmetric spin paths,
\begin{equation}\label{sojblip}
\fl\qquad
\begin{array}{rcl}
\xi_1^{}(t') &=& {\displaystyle \frac{1}{2}[\sigma(t')-\sigma'(t')]\; , \hspace{1cm}
\eta_1^{}(t')\;=\; \frac{1}{2}[\sigma(t')+\sigma'(t')] }\;,\\[3mm]
\xi_2^{}(t') &=& {\displaystyle \frac{1}{2}[\tau(t')-\tau'(t')]\; , \hspace{1cm}
\eta_2(^{}t')\;=\;\frac{1}{2}[\tau(t')+\tau'(t')] }\; .
\end{array}
\end{equation}
The correlator $Q_\zeta^{}(t)\,=\,Q'_\zeta(t)\,+\,{\rm i}\,Q''_\zeta(t)$ is the second integral of the force auto\-correlation 
function $\langle X_\zeta^{}(t) X_\zeta^{}(0)\rangle_\beta$ (see eq.~(\ref{powerspec})). In the Ohmic scaling limit, 
we  have
\begin{equation}\label{ohmcorr}
\fl\quad
Q_\zeta(t)\;=\; 2K_\zeta \,\left[\ln \left(\frac{\beta \omega_{\rm c}^{}}{\pi} \right)
+\ln \sinh \left( \frac{\pi |t|}{\beta} \right)   \right] 
+ i \pi K_\zeta \, {\rm sgn}(t)\;, \qquad \zeta=1,\,2\; .
\end{equation} 
Here, $K_\zeta$ is the usual dimensionless Ohmic coupling strength for the spin $\zeta$, and  
$\beta = 1/T$ is the inverse temperature.

To handle the quadruple path integral (\ref{rdm}), we follow the procedure for the single spin-boson problem 
\cite{leggett,bookweiss} and write it as an integral over two paths, one for each spin. Each such path visits
the diagonal "sojourn" states and the off-diagonal "blip" states of the respective spin. 
A path which starts and ends in a sojourn state
must contain an even number of transitions with amplitude $\mp{\rm i}\,\Delta_{\zeta}/2$ for each flip of 
spin $\zeta$. The flips occur at times $t_j^{}$ for spin 1 and at times $s_j^{}$ for spin 2. Upon
labeling the sojourn and blip states with charges $\eta_{1,j}^{},\,\xi_{1,j}^{}$ (for spin 1) and
$\eta_{2,j}^{},\,\xi_{2,j}^{}$ (for spin 2), each with values $\pm 1$, the paths with $2n_1^{}$ and $2n_2^{}$ transitions,
respectively,  may be written as
\begin{eqnarray}
\eta^{(n_1^{})}_1(t') &=& \sum_{j=0}^{n_1^{}} \eta_{1,j}^{} \left[\theta(t'-t_{2j})-\theta(t'-t_{2j+1}) \right]\; , \\
\xi^{(n_1^{})}_1(t') &=&  \sum_{j=1}^{n_1^{}} \xi_{1,j}^{}  \left[\theta(t'-t_{2j-1})-\theta(t'-t_{2j}) \right]\; ,  \\
\eta^{(n_2^{})}_2(t') &=& \sum_{j=0}^{n_2^{}} \eta_{2,j}^{} \left[\theta(t'-s_{2j}^{})-\theta(t'-s_{2j+1}^{}) \right]\; , \\
\xi^{(n_2^{})}_2(t')  &=& \sum_{j=1}^{n_2^{}} \xi_{2,j}^{}  \left[\theta(t'-s_{2j-1}^{})-\theta(t'-s_{2j}^{}) \right]\; .
 \end{eqnarray}
Upon introducing the notation $Q_{\zeta;j,k} = Q_\zeta(t_j-t_k)$, we may write the bath correlations between the blip pair
$\{j,k\}$ of spin $\zeta$ in the compact form
\begin{equation}\label{blipcor}
\fl\qquad
\Lambda_{\zeta;j,k} \;=\; Q'_{\zeta;2j,2k-1}\,+\,Q'_{\zeta;2j-1,2k}\,-\,Q'_{\zeta;2j,2k}\,-\,Q'_{\zeta;2j-1,2k-1} \; .
\end{equation}
With this, the influence functional for the paths (\ref{sojblip}) reads
\begin{equation}\label{infl}
\begin{array}{rcl}
\fl
{\cal F}_\zeta^{(n_\zeta^{})} &=& {\displaystyle \exp\Big[- \sum_{j=1}^{n_\zeta^{}}Q'_{\zeta;2j,2j-1}\Big]\,
 \exp\Big[-\sum_{j=2}^{n_\zeta^{}}\sum_{k=1}^{j-1}\xi_{\zeta j}^{}\xi_{\zeta,k}^{}\Lambda_{\zeta;j,k}^{}\Big] } \\[4mm]
 && \times\; {\displaystyle 
 \,\exp\Big[{\rm i} \,\pi K_\zeta^{}\sum_{k=0}^{n_\zeta^{}-1} \xi_{\zeta,k+1}^{}\eta_{\zeta,k}^{}    \Big] } \; .
 \end{array}
\end{equation}
Here the first and second term represent the intrablip and interblip correlations, respectively. 
The phase term  is specific to the Ohmic scaling limit and represents
correlations of the sojourns with their subsequent blips.

The sum over all paths now means (i) to sum over all possible intermediate sojourn and blip states of the two spins
the paths with a given number of transitions can visit, (ii) to integrate over the (for each spin) time-ordered 
jumps of these paths, and (iii) to sum over the possible number of transitions the two spins can take,
\begin{equation}\label{pathsum}
\fl\;\;
\begin{array}{rcl}
{\displaystyle \sum_{\rm all\; paths}\!\!\! \cdots }&\to & 
{\displaystyle \sum_{n_1^{}=0}^\infty 
\sum_{ \{\eta_{1,j}^{}=\pm 1\} } 
\sum_{ \{\xi_{1,j}^{}=\pm 1\} } 
\int_{t_{\rm i}^{}}^{t_{\rm f}^{}}\!\! {\rm d}t_{2n_1^{}}^{}
\int_{t_{\rm i}^{}}^{t_{2n_1^{}}} \!\!  {\rm d}t_{2n_1^{}-1}^{}  \cdots 
\int_{t_{\rm i}^{}}^{t_2^{}} 
\!\! {\rm d}t_1^{}    }           \\[6mm]
&\times& {\displaystyle \sum_{n_2^{}=0}^\infty 
\sum_{ \{\eta_{2,j}^{}=\pm 1\} } 
\sum_{ \{\xi_{2,j}^{}=\pm 1\} } 
\int_{t_{\rm i}^{}}^{t_{\rm f}^{}} \!\! {\rm d}s_{2n_2^{}}^{}
\int_{t_{\rm i}^{}}^{s_{2n_2^{}}} \!\!  {\rm d}s_{2n_2^{}-1}^{}  \cdots 
\int_{t_{\rm i}^{}}^{s_2^{}} 
\!\!{\rm d}s_1^{}\;\cdots   }   \; .
\end{array}
\end{equation}
For coherences, the number of transitions in the respective spin path is odd.

Next, we rewrite the double path sum (\ref{pathsum}) with time-ordering for each spin in terms of a single path 
over the 16 possible states with time ordering of all the flip times $\{t_j^{}\},\,\{s_i^{}\}$. 
In this representation, the system lingers for some period in a particular state of the double-spin RDM and then it flips with
amplitude $\pm\,\rmi\,\Delta_1^{}/2$ or $\pm\,\rmi\,\Delta_2^{}/2$ to another state.
For the longitudinal spin-spin coupling $-\frac{1}{2} v \sigma_z^{}\tau_z^{}$ in (\ref{ham1}), the coupling factor 
${\cal B}$ in eq.~(\ref{rdm})
is unity when the system dwells on one of the four pure soujourn or one of the four pure blip states, and it is 
sensitive to the coupling when it stays, say for a period $u_j^{}$, in one of the eight hybrid states. In detail, we have
\begin{equation}\label{intfac}
\fl\qquad
\begin{array}{rcl}
{\cal B}[\,j\,] &=& {\rm e}^{+i v u_j^{}}_{}\;, \quad\quad\mbox{for}\quad\quad j \;\in\; (1,2),\,(2,1),\,(4,3),\,(3,4)\; , \\[3mm]
{\cal B}[\,j\,] &=& {\rm e}^{-i v u_j^{}}_{}\;, \quad\quad\mbox{for}\quad\quad j \;\in\; (1,4),\,(4,1),\,(2,3),\,(3,2)\; . 
\end{array}
\end{equation}

Combination of the above expressions yields the exact formal solution for the dynamics of the RDM of the two-spin system 
in the Ohmic scaling limit. Evidently, because of the nonconvolutive form of the bath correlations in the influence 
functional (\ref{infl}), the path sum can not be performed in analytic form. Alternatively, one may recast the exact 
formal series expression for the populations in the form of generalized master equations  in which the kernels, 
by definition, are the irreducible components of path segments with diagonal initial and final states \cite{bookweiss}.
In the general case, the kernels are given by an infinite series in $\Delta_1^{}$ and $\Delta_2^{}$, 
with the time integrals in each summand being again in nonconvolutive form. 
Additional difficulties in performing the path sum (\ref{pathsum}) arises from the spin-spin coupling (\ref{intfac}).

\section{Exact solution in analytic form in the white-noise limit}

In the white-noise limit of eq.~(\ref{powerspec}), $S_\zeta^{}(\omega\ll 1/\beta) \,=\, 2\vartheta_\zeta^{} $, where
\begin{equation}\label{whitenoise}
\vartheta_\zeta^{} \;=\; 2\pi K_\zeta^{} T  
\end{equation}
is a scaled thermal energy,
the bath  correlation function takes the form\footnote{The term $\Upsilon_\zeta $ accounts for deviation of the
actual high frequency behavior of $G_\zeta(\omega)$  from the exponential cut-off form in eq.~(\ref{specdens}) \cite{bookweiss}.} 
\begin{equation}\label{wncorr}
Q_\zeta(t)=2K_\zeta\Big[\, \ln \Big(\frac{\omega_{\rm c}^{}}{2 \pi T}\Big)\,+\, \Upsilon_\zeta \,\Big] \,+\, \vartheta_\zeta^{} |t|
 \,+\, {\rm i}\, \pi K_\zeta \, {\rm sgn}(t) \;.
\end{equation} 
This expression emerges directly from eq. (\ref{ohmcorr}) in the high temperature or long-time limit $t/\beta\gg 1$.
The first term in eq.~(\ref{wncorr}) leads to an adiabatic (Franck-Condon-type) renormalization factor made up 
by modes in the frequency range $2\pi T <\omega < \omega_{\rm c}^{}$. It is natural to assimilate  this term, together 
with the phase term, into an effective temperature-dependent tunneling matrix element,
\begin{equation}\label{adia}
\fl\quad
\bar\Delta_\zeta^{2} \;=\;\Delta_{\zeta,T}^{2} \cos(\pi K_\zeta^{})\; ,\qquad\mbox{with}\qquad 
\Delta_{\zeta,T}^{2} \;=\; (2\pi T/\Delta_{\zeta,\rm r}^{})^{2K_\zeta^{}}_{}\,\Delta_{\zeta,\rm r}^{2}\,
\rme^{-2K_\zeta\Upsilon_\zeta}_{} \;,
\end{equation}
where $\Delta_{\zeta,\rm r}^{1- K_\zeta^{}}\,=\, \Delta_\zeta^{}/\omega_{\rm c}^{K_\zeta^{}}$ is the standard renormalized
tunneling matrix element. 

All dynamical effects of the environmental coupling are captured by the second term $ \vartheta_\zeta^{}|t|$ 
in eq.~(\ref{wncorr}). From this we see that the weight of the thermal energy relative to the systems energies 
$\bar\Delta_\zeta^{}$ and $v$ is assessed by the scaled thermal energy $\vartheta_\zeta^{}$.

Based on experience from the single spin-boson system we should expect that the form  (\ref{wncorr}) is a remarkably 
good approximation in the parameter range \cite{bookweiss}
\begin{equation}\label{tempreg}
\fl\qquad K_\zeta^{} \,\lapx \, 0.3\; , \qquad\mbox{and}\qquad \Omega_\pm^{} \; \lapx \; T \;\ll \omega_{\rm c}^{}  \; ,
\end{equation} 
and this is corroborated indeed by our study.
For a single unbiased spin with Ohmic damping $K\ll 1$, the coherent-incoherent "phase"-transition is at temperature
$T = T^\ast_{}$ with $T^\ast_{} = \Delta_{\rm r}^{}/(\pi K)$ \cite{bookweiss}. Therefore we should expect, that, 
for $K_\zeta^{}\ll 1$, the white noise form (\ref{wncorr}) is valid not only in the incoherent regime but also in a sizeable 
domain of the coherent regime. This shall be confirmed subsequently.

For the form (\ref{wncorr}) of $Q_\zeta^{}(t)$, the interblip correlations cancel out exactly in eq.~(\ref{blipcor}),
$\Lambda_{\zeta;j,k}^{}=0$. As a result, each term of the infinite series for the RDM becomes a convo\-lution. This makes
the path sum accomplishable.
We shall now exemplify, by taking $\langle\sigma_z^{}\rangle_t$ and $\langle\sigma_z^{}\tau_z^{}\rangle_t$
as examples, that the path sum for the Laplace transform of the RDM can be carried out 
exactly in analytic form. This is achieved by first calculating the kernels and then summing up 
the respective geometrical series of these objects. 

\subsection{The expectations $\langle \sigma_z^{}\rangle_t$ and  $\langle \tau_z^{}\rangle_t$}

\begin{figure}[h!]
\vspace{0cm}
\begin{center}
\includegraphics[scale=.30]{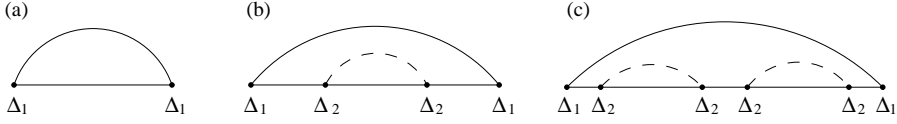}
\end{center}
\caption{\label{f1}\small Sketch of the irreducible kernels (with amputated legs)
 ${\cal K}_{1,0}^{}(\lambda)$ (a), ${\cal K}_{1,1}^{}(\lambda)$ (b), and  ${\cal K}_{1,2}^{}(\lambda)$ (c). 
The solid and dashed curve represent the correlations of bath 1 and of bath 2, respectively.}
\end{figure}

By definition, the kernels represent irreducible path segments which interpolate between pure sojourn states. 
Irreducibility means that these segments can not be separated 
into uncorrelated pieces without removing bath correlations. The analysis gives that every contribution to the kernel 
${\cal K}(\lambda)$ of $\langle\sigma_z^{}(\lambda)\rangle$ displays initially and finally a transition of spin $\sigma$ 
with any number of even hops of spin $\tau$ at intermediate times, as shown in the diagrams of 
Fig.~\ref{f1}.  There are no other contributions.
For all other irreducible diagrams, one may think of, e.g., where either the first or the last flip, 
or both of them, are flips of spin $\tau$, the respective contributions from the two different final states of spin $\tau$ 
cancel each other.

It is convenient to write 
${\cal K}(\lambda)\,=\, \sum_{n=0}^\infty  {\cal K}_{1,n}^{}(\lambda)$, where $ {\cal K}_{1,n}^{}(\lambda)$ is of order
$\Delta_1^2\,\Delta_2^{2n}$. The sum over all spin states of order $(1,n)$ yields the expressions
\begin{equation}\label{k1}
\fl\quad
{\cal K}_{1,0}^{} (\lambda) \;=\; - \,\Delta_{1,T}^2 \,\int_0^\infty \!\!  \rmd\tau_1^{}\,
{\rm e}^{-(\lambda+\vartheta_1^{})\,\tau_1^{}}_{} \, \cos(\pi\,K_1^{}\,+\, v\,\tau_1^{}) \; ,
\end{equation}
\begin{equation}
\fl\quad
\begin{array}{rcl}
 {\cal  K}_{1,1}^{} (\lambda)  &=& {\displaystyle \Delta_{1,T}^2\,\Delta_{2,T}^2  \,
\int_0^\infty \!\! \rmd\tau_1^{}\, \rmd\tau_2^{}\,\rmd\tau_3^{}\,
{\rm e}^{-(\lambda+\vartheta_1^{})(\tau_1^{}+\tau_2^{}+\tau_3^{})}_{}\, {\rm e}^{-\vartheta_2^{}\,\tau_2^{}}_{} } \\[3mm]
&&\!\!\!\!\times \; {\displaystyle  \,\frac{1}{2}\,[\, \cos(\pi K_1^{} + v\tau_1^{} + v \tau_3^{}) -  
\cos(\pi K_1^{} + v\tau_1^{} - v \tau_3^{})\,] \cos(\pi\,K_2^{})} \; ,
\end{array}
\end{equation}
\begin{equation}
\fl\quad
{\cal K}_{1,n}^{}(\lambda) \;=\; {\cal K}_{1,1}^{}(\lambda) \left( -\,\bar\Delta_{2}^2 \int_0^\infty\!\!\rmd \tau_1^{}\,
\rmd \tau_2^{}\,  {\rme}^{-(\lambda+\vartheta_1^{})\tau_1^{}}_{} \,
\rme^{-(\lambda +\vartheta_1^{}+\vartheta_2^{})\tau_2^{}}_{}\,\cos(v\tau_1^{})\right)^{n-1}  \; .
\end{equation}
Here we have taken into account that, for the white-noise form (\ref{wncorr}), a correlation of bath $\zeta$ 
stretching over an interval between neighboring hops effectively leads to a shift of the Laplace variable in the 
respective time integral, $\lambda \to\lambda+\vartheta_\zeta^{}$. It is convenient to split the kernels into the
contributions which are even and odd in the coupling $v$, 
${\cal K}(\lambda) = {\cal K}^{(+)}_{}(\lambda)+ {\cal K}^{(-)}_{}(\lambda)$. The resulting expressions may 
be written as
\begin{equation}\label{kplus}
\fl\quad
\begin{array}{rcl}
{\cal K}^{(+)}_{}(\lambda)\!\! &=&\!\! {\displaystyle - \bar\Delta_{1}^2 \,
\frac{\lambda + \vartheta_1^{}}{(\lambda +\vartheta_1^{})^2_{}+ v^2_{}} } \\[4mm]
&\times & {\displaystyle 
\!\!\!\left\{ 1 \;+\; \frac{v^2_{}}{(\lambda +\vartheta_1^{})(\lambda +\vartheta_1^{}+\vartheta_2^{})}\,      
\frac{\bar\Delta_{2}^2 }{  ( \lambda +\vartheta_1^{})^2_{}+ v^2_{}   +  
\frac{\lambda+\vartheta_1^{}}{\lambda +\vartheta_1^{} + \vartheta_2^{}}\bar\Delta_2^{2} }\right\}  }\; ,
\end{array}
\end{equation}
\begin{equation}\label{kminus}
\fl\quad
\begin{array}{rcl}
{\cal K}^{(-)}_{}(\lambda)\!\! &=&\!\! {\displaystyle  \tan(\pi K_1^{})\,\bar\Delta_{1}^2\,
\frac{v}{(\lambda +\vartheta_1^{})^2_{}+ v^2_{}} } \\[4mm]
&\times & {\displaystyle 
\!\!\!\left\{ 1 \;-\; \frac{\lambda+\vartheta_1^{}}{\lambda +\vartheta_1^{}+\vartheta_2^{}}\,      
\frac{\bar\Delta_{2}^2 }{  ( \lambda +\vartheta_1^{})^2_{}+ v^2_{}   +  
\frac{\lambda+\vartheta_1^{}}{\lambda +\vartheta_1^{} + \vartheta_2^{}}\bar\Delta_2^{2} }\right\}  } \; .
\end{array}
\end{equation}
Paths which visit a pure sojourn state at intermediate times yield reducible con\-tri\-butions.
Taking into account all possibilities of such visits yields a geometrical series in the kernel
${\cal K}^{(+)}_{}(\lambda)$, while ${\cal K}^{(-)}_{}(\lambda)$ occurs only once as initial irreducible
contribution. Thus we get for $\langle\sigma_z^{}(\lambda)\rangle$ the concise form
\begin{equation}\label{expecsigma}
\fl
\langle\sigma_z^{}(\lambda) \rangle \;=\; \frac{1}{\lambda}\,
\frac{ 1\,+\, {\cal K}^{(-)}_{}(\lambda)/\lambda }{ 1 \,-\, {\cal K}^{(+)}_{}(\lambda)/\lambda } 
\;=\; \frac{1}{\lambda}\,\frac{N_1^{}(\lambda)}{D_1^{}(\lambda)} \;.
\end{equation}
Algebraic manipulation gives $\langle\sigma_z^{}(\lambda)\rangle$ in the form of a simple fraction with denominator and
numerator in the form of polynomials of degree four. We get
\begin{equation}\label{sz1}
\fl
 \begin{array}{rcl}
N_1^{}(\lambda) &=& \lambda^4_{} \,+\, (3\, \vartheta_1^{} + \vartheta_2^{})\lambda^3_{} \,+\, 
(v_{}^2 + \bar\Delta_{2}^2 + 3\, \vartheta_1^2 + 2\, \vartheta_1^{}\, \vartheta_2^{})\lambda^2_{} \\[2mm]
&+& \!\!\!(v^2_{}\vartheta_1^{} + \bar\Delta_2^2\vartheta_1^{} + \vartheta_1^3 + v^2_{}\vartheta_2^{} 
+\vartheta_1^2\vartheta_2^{})\lambda + v\bar\Delta_1^{}\tan(\pi K_1^{}) (\vartheta_1^{}+\vartheta_2^{} +\lambda)  \; , 
\end{array}
\end{equation}
\begin{equation}\label{sz2}
\fl
\begin{array}{rcl}
D_1^{}(\lambda) &=& \lambda^4_{} \,+\, (3\, \vartheta_1^{} + \vartheta_2^{})\,\lambda^3_{} \,+\, (\bar\Omega_{}^2 + 
\bar\delta_{}^2 + 3\, \vartheta_1^2 + 2\, \vartheta_1^{}\, \vartheta_2^{})\,\lambda^2_{}  \\[3mm]
&+& (v^2\, \vartheta_1^{} + 2\, \bar\Delta_{1}^2 \,\vartheta_1^{} + \bar\Delta_{2}^2 \, \vartheta_1^{} + 
\vartheta_1^3 + 
  v^2_{} \,\vartheta_2^{} + \bar\Delta_{1}^2\, \vartheta_2^{} + \vartheta_1^2\, \vartheta_2^{})\,\lambda  \\[3mm]
&+& \bar\Delta_{1}^2 \,\vartheta_1^2 + \bar\Delta_{1}^2\, \vartheta_1^{}\, \vartheta_2^{} + 
\bar\Omega_{}^2 \,\bar\delta_{}^2  \; .
\end{array}
\end{equation}
Here we have introduced the eigenfrequencies $\Omega$, $\delta$ of the undamped coupled two-spin system. 
The bar denotes adiabatic renormalization in eq.~(\ref{def_eigen1}) according to eq.~(\ref{adia}).  

The pole at $\lambda=0$ in  eq.~(\ref{expecsigma}) yields the equilibrium value 
\begin{equation}\label{sigzetequ}
\langle\sigma_z^{}\rangle_{\rm eq}^{}\;=\; - \frac{{\cal K}^{(-)}_{}(0)}{{\cal K}^{(+)}_{}(0)} 
\;=\; \frac{v}{2T}\,\frac{ \vartheta_1^{}(\vartheta_1^{}+\vartheta_2^{})}{
\bar\Delta^2_2 \,+\, \vartheta_1^{}(\vartheta_1^{}+\vartheta_2^{})} \; .
\end{equation}
Hence $\langle\sigma_z^{}\rangle_{\rm eq}^{}$ is negligibly small for $\bar\Delta_2^{}\neq 0$, while, as $\bar\Delta_2^{} \to 0$, 
it takes the proper (white noise) equilibrium value $v/(2T)$ of the single biased spin boson system.

The dynamical poles $\lambda_j^{}$ ($j=1,\cdots,\,4$) are given by a quartic equation with real coefficients. 
Upon collecting the various pole contributions, we get in the time domain
\begin{equation}\label{sztime}
\langle\sigma_z^{}\rangle_t^{} \;=\; \sum_{i=1}^4 A_i^{}\,\rme^{\lambda_i^{}t}_{} \,+\, \langle\sigma_z^{}\rangle_{\rm eq}^{} \; .
\end{equation}
Evidently, the expectation $\langle\tau_z^{}(\lambda) \rangle$ takes similar form,
\begin{equation}\label{expectau}
\fl\qquad
\langle\tau_z^{}(\lambda) \rangle 
\;=\; \frac{1}{\lambda}\,\frac{N_2^{}(\lambda)}{D_2^{}(\lambda)} \; .
\end{equation}
Here, the polynomials $N_2(\lambda)$ and $D_2(\lambda)$ follow from the expressions (\ref{sz1}) and (\ref{sz2})  by 
interchange of the indices 1 and 2.
\begin{figure}[h!]
\vspace{0cm}
\begin{center}
\includegraphics[scale=.33]{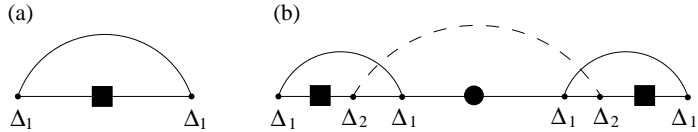}
\end{center}
\caption{\label{f2}\small Sketch of the irreducible kernels ${\cal A}_1^{\rm (a)}(\lambda)$ (left)
and  ${\cal A}_1^{\rm (b)}(\lambda)$ (right). The intervals are dressed by self-energy contributions of spin $\sigma$ 
(black circle) and spin $\tau$ (black square), as sketched in Fig.~\ref{f4}.}
\end{figure}
\begin{figure}[h!]
\vspace{0cm}
\begin{center}
\includegraphics[scale=.33]{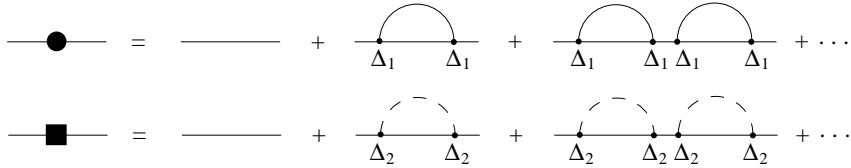}
\end{center}
\caption{\label{f4}\small Self-energy terms due to spin $\sigma$ (circle) and spin $\tau$ (square). }
\end{figure}

\subsection{The expectation $\langle \sigma_z^{}\tau_z^{}\rangle_t$}
\begin{figure}[h]
\vspace{0cm}
\begin{center}
\includegraphics[scale=.33]{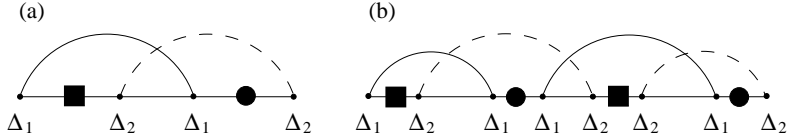}
\end{center}
\caption{\label{f3}\small Sketch of the irreducible kernels ${\cal B}_1^{\rm (a)}(\lambda)$ (left)
and  ${\cal B}_1^{\rm (b)}(\lambda)$ (right). Again, the intervals are dressed as sketched in Fig.~\ref{f4}.}
\end{figure}

Consider first contributions to the kernel of $\langle\sigma_z^{}\tau_z^{}(\lambda)\rangle$ in which the first and last flip 
are made by spin $\sigma$, as sketched in Fig.~\ref{f2}. In the intervals of the bare diagrams, either spin $\sigma$ or spin
$\tau$ or both stay offdiagonal. Every interval in which spin $\zeta$ dwells in a sojourn state is dressed by
selfenergy contributions schematically given for spin $\sigma$ (circle) and spin $\tau$ (square) in Fig.~\ref{f4}.
Diagram \ref{f2}\,(a) yields
\begin{eqnarray*}
\fl
{\cal A}_{1}^{\rm (a)}(\lambda) &=& {\cal K}_{1,0}^{}(\lambda)
\sum_{n=0}^\infty  \left( -\,\bar\Delta_{2}^2 \int_0^\infty\!\!\rmd \tau_1^{}\,
\rmd \tau_2^{}\,  {\rme}^{-(\lambda+\vartheta_1^{})\tau_1^{}}_{} \,
\rme^{-(\lambda +\vartheta_1^{}+\vartheta_2^{})\tau_2^{}}_{}\,\cos(v\tau_1^{})\right)^n \\
\fl
&=& -\,(\lambda +\vartheta_1^{}+\vartheta_2^{})\Big[\,1 \,-\, \frac{v}{\lambda+\vartheta_1^{}}\tan(\pi K_1^{})\,\Big]
\,\alpha_1^{}(\lambda)  \; ,
\end{eqnarray*}
where the functions $\alpha_\zeta^{}$ $(\zeta =1,\,2)$ are given by
\begin{eqnarray}\label{alph1}
\fl\qquad
\alpha_1^{}(\lambda) &=& \frac{\lambda +\vartheta_1^{}}{\lambda+\vartheta_1^{}+\vartheta_2^{}} \,
\frac{\bar\Delta_1^2}{(\lambda+\vartheta_1^{})^2+ v^2_{} + \frac{\lambda +\vartheta_1^{}}{\lambda+
\vartheta_1^{}+\vartheta_2{}}\bar\Delta_2^2}  \; , \\  \label{alph2}
\fl\qquad
\alpha_2^{}(\lambda) &=& \frac{\lambda +\vartheta_2^{}}{\lambda+\vartheta_1^{}+\vartheta_2^{}} \,
\frac{\bar\Delta_2^2}{(\lambda+\vartheta_2^{})^2+ v^2_{} + \frac{\lambda +\vartheta_2^{}}{\lambda+
\vartheta_1^{}+\vartheta_2{}}\bar\Delta_1^2} \; .
\end{eqnarray}
The nested diagram in Fig.~\ref{f2}\,(b) produces the additional factor $\alpha_2^{}(\lambda)\,\alpha_1^{}(\lambda)$,
\begin{equation}
\fl\qquad
{\cal A}_1^{\rm (b)}(\lambda) \;=\;  {\cal A}_{1}^{\rm (a)}(\lambda)\, \alpha_2^{}(\lambda)\,\alpha_1^{}(\lambda)  \; .
\end{equation}
Higher-order nested diagrams of the type sketched in Fig.~\ref{f2} class into a geometrical series. All these
terms are readily summed up to the contribution
\begin{equation}\label{contriba1}
\fl\qquad
{\cal A}_1^{}(\lambda) \;=\; {\cal A}_{1}^{\rm (a)}(\lambda)\,
\,\frac{1}{1 \,-\, \alpha_2^{}(\lambda)\alpha_1^{}(\lambda)}   \; .
\end{equation}
Similarly, we find that diagrams (a) and (b) in Fig.~\ref{f3} yield the expressions
\begin{equation}
\fl\qquad
{\cal B}_1^{\rm (a)}(\lambda) \;=\; \,(\lambda +\vartheta_1^{}+\vartheta_2^{})
\Big[\,1 \,-\, \frac{v}{\lambda+\vartheta_1^{}}\tan(\pi K_1^{})\,\Big]
\,\alpha_1^{}(\lambda)\alpha_2^{}(\lambda)  \; ,
\end{equation}
\begin{equation}
\fl\qquad
{\cal B}_1^{\rm (b)}(\lambda) \;=\; {\cal B}_1^{\rm (a)}(\lambda)\,\alpha_1^{}(\lambda)\,\alpha_2^{}(\lambda)  \; .
\end{equation}
With all higher-order nested diagrams of this type added, one finds again a geometrical series in 
$\alpha_1^{}(\lambda)\,\alpha_2^{}(\lambda)$,
\begin{equation}\label{contribb1}
\fl\qquad
{\cal B}_1^{}(\lambda) \;=\; {\cal B}_1^{\rm (a)}(\lambda)
\,\frac{1}{1\,-\, \alpha_1^{}(\lambda)\alpha_2^{}(\lambda)}  \; .
\end{equation}
Clearly, we must add those contributions resulting from the terms ${\cal A}_1^{}(\lambda)$
and ${\cal B}_1^{}(\lambda)$ by interchange of the spins $\sigma$ and $\tau$. The analysis shows that there are no other
con\-tri\-bu\-tions. Again, we split the kernel ${\cal C}(\lambda)$ into the parts which are even and odd in the coupling $v$. 
We readily get for
${\cal C}^{(\pm)}_{}(\lambda) = \sum_{\zeta=1,2}[\,{\cal A}^{(\pm)}_{\zeta}(\lambda) + {\cal B}^{(\pm)}_{\zeta}(\lambda) \,]$
the forms
\begin{equation}\label{cplus}
\fl\qquad
{\cal C}^{(+)}_{}(\lambda)\;=\; -\,(\lambda +\vartheta_1^{}+\vartheta_2^{})\,
\frac{\alpha_1^{}(\lambda)\, + \,\alpha_2^{}(\lambda) \,-\,2\alpha_1^{}(\lambda) \alpha_2^{}(\lambda) }{
1 \,-\, \alpha_1^{}(\lambda) \alpha_2^{}(\lambda)}  \; ,
\end{equation}
\begin{equation}\label{cminus}
\fl\qquad
{\cal C}^{(-)}_{}(\lambda)\;=\; ( \lambda +\vartheta_1^{}+\vartheta_2^{} )\,
\sum_{\zeta=1,2} \frac{v}{\lambda+\vartheta_\zeta^{}}\tan(\pi K_\zeta^{}) \frac{  
 \alpha_\zeta^{}(\lambda) \,-\,  \alpha_1^{}(\lambda)\alpha_2^{}(\lambda)}{ 1 \,-\, \alpha_1^{}(\lambda) \alpha_2^{}(\lambda)} \; .   
\end{equation}
These expression represent the entity of irreducible path segments.
Next, we observe that the sum of two-spin paths with any number of interim visits of pure sojourn states yields a geometrical series
of these objects. In the part which is odd in $v$, the first irreducible path section is again described by
the kernel ${\cal C}^{(-)}(\lambda)$. Thus we get
\begin{equation}\label{expecsigtau}
\fl\qquad
\langle\sigma_z^{}\tau_z^{}(\lambda) \rangle \;=\; \frac{1}{\lambda}\,
\frac{ 1\,+\, {\cal C}^{(-)}_{}(\lambda) /\lambda }{
1 \,-\, {\cal C}^{(+)}_{}(\lambda)/\lambda} \;=\;  \frac{1}{\lambda}\,\frac{N_{}^{}(\lambda)}{D_{}^{}(\lambda)} \; .
\end{equation}
The second form is a simple fraction with polynomials $N(\lambda)$ and $D(\lambda)$ of sixth order,
\begin{equation}\label{sigtauz2}
\fl
\begin{array}{rcl}
N(\lambda)&=& 
v^3_{}(\vartheta_1^{}+\vartheta_2^{}+\lambda )[\,\tan(\pi K_1^{})\, \bar\Delta_1^2 \,+\, \tan(\pi K_2^{})\, \bar\Delta_2^2 \,]  \\[2mm]
&& +\; v \tan( \pi K_1^{} )\, \bar\Delta_1^2\,(\vartheta_2^{}+\lambda)\,[\,\bar\Delta_1^2\, -\, \bar\Delta_2^2\, +\,
(\vartheta_2+\lambda)(\vartheta_1^{}+\vartheta_2^{}+\lambda)  \,]   
  \\[2mm]
 && + \;v \tan( \pi K_2^{} )\, \bar\Delta_2^2\,(\vartheta_1^{}+\lambda)\,[\,\bar\Delta_2^2 \,-\, \bar\Delta_1^2 \,+\,
(\vartheta_1+\lambda)
(\vartheta_1^{}+\vartheta_2^{}+\lambda)  \,]   
  \\[2mm] 
  &&+\; \lambda \, v^4_{}(\vartheta_1^{}+\vartheta_2^{}+\lambda) 
\,+\,\lambda v^2_{}[\,\bar\Delta_2^2\,(\vartheta_1^{}+\lambda) +  \bar\Delta_1^2\,(\vartheta_2^{}+\lambda)\,] \\[2mm]
&& + \; \lambda \,v^2_{} 
(\vartheta_1^{}+\vartheta_2^{}+\lambda) [\,\vartheta_1^2+\vartheta_2^2 +2\lambda (\vartheta_1^{}+\vartheta_2^{}+\lambda)\,]\\[2mm]
&&+ \; \lambda \,(\vartheta_1^{}+\lambda)(\vartheta_2^{}+\lambda)
[\,\bar\Delta_1^2\,(\vartheta_1+\lambda) + \bar\Delta_2^2\,(\vartheta_2+\lambda)\,] \\[2mm]
&& + \; \lambda\,(\vartheta_1^{}+\lambda)^2_{}(\vartheta_2^{}+\lambda)^2_{}(\vartheta_1^{}+\vartheta_2^{}+\lambda) \; ,
\end{array}
\end{equation}
\begin{equation}\label{sigtauz1}
\fl
\begin{array}{rcl}
D(\lambda)&=& \lambda^6_{} \,+ \, 3(\vartheta_1 +\vartheta_2)\,\lambda^5_{} \,+\,
(\bar\Omega_+^2 + \bar\Omega_-^2 + 3\vartheta_1^2 + 8 \vartheta_1 \vartheta_2 + 3 \vartheta_2^2) \,\lambda^4_{}  \\[2mm]
&&+\;(\vartheta_1 +\vartheta_2)(2\bar\Omega_+^2 + 2 \bar\Omega_-^2 +\vartheta_1^2 + 6\vartheta_1\vartheta_2 +
\vartheta_2^2)\,\lambda^3_{} \\[2mm]
&&+ \;[\,\bar\Omega_+^2\,\bar\Omega_-^2 + \vartheta_1 \vartheta_2(2 \vartheta_1 +\vartheta_2)(\vartheta_1+2\vartheta_2)
+ v^2 (3\vartheta_1^2 + 4\vartheta_1\vartheta_2 +3\vartheta_2^2) \\[2mm]
&& +\; \bar\Delta_1^2(2\vartheta_1 +\vartheta_2)(\vartheta_1 +3\vartheta_2) +
   \bar\Delta_2^2(2\vartheta_2 +\vartheta_1)(\vartheta_2 +3\vartheta_1) ]\,\lambda_{}^2\\[2mm]
&& +\;(\vartheta_1 +\vartheta_2)[\,\bar\Omega_+^2\, \bar\Omega_-^2  +\vartheta_1^2\vartheta_2^2 
+ (v^2+ \bar\Delta_2^2)\vartheta_1^2 + (v^2_{}+\bar\Delta_1^2)\vartheta_2^2 \\[2mm]
&& +\; 3\vartheta_1^{}\vartheta_2^{}(\bar\Delta_1^2+\bar\Delta_2^2) \,]\,\lambda 
\;+ \; v^2(\vartheta_1 + \vartheta_2)(\bar\Delta_1^2\vartheta_1 + \bar\Delta_2^2 \vartheta_2)\\[2mm]
&& +\;  \vartheta_1\vartheta_2 [\, (\bar\Delta_1^2 - \bar\Delta_2^2)^2 
+ (\bar\Delta_2^2\vartheta_1 + \bar\Delta_1^2\vartheta_2 )(\vartheta_1^{}+\vartheta_2^{})  \,] \; .
\end{array}
\end{equation}

The odd powers  of the pole function can be removed with a shift. 
Putting $D[\,\lambda=x-(\vartheta_1^{}+\vartheta_2^{})/2\,] \equiv \tilde D(x)$, 
we obtain a polynomial with even powers,
\begin{equation}\label{sigtauz3}
\fl
\begin{array}{rcl}
\tilde D(x) &=& x^6_{} \,+\, [\,\bar\Omega_+^2+\bar\Omega_-^2 - \frac{1}{2}(\vartheta_1^2+\vartheta_2^2)
-\frac{1}{4} (\vartheta_1^{}-\vartheta_2^{})^2_{}\,]\,x^4_{} \\[3mm]
&&+\;[\,\bar\Omega_+^2\bar\Omega_-^2\, +\,(\bar\Delta_1^2+\bar\Delta_2^2-2v^2_{})\vartheta_1^{} \vartheta_2^{}
-\bar\Delta_1^2\vartheta_1^2-\bar\Delta_2^2\vartheta_2^2 \\[3mm]
&&\quad\; +\;\frac{1}{16}  (\vartheta_1^{}-\vartheta_2^{})^2_{}(3\vartheta_1^2+2\vartheta_1^{}\vartheta_2^{}+3\vartheta_2^2 ) \,]\,x^2_{} \\[3mm]
&&-\frac{1}{4} [\,v^2_{}(\vartheta_1^{}+\vartheta_2^{}) \,+\, (\bar\Delta_1^2-\bar\Delta_2^2)(\vartheta_2^{}-\vartheta_1^{})
\,+\,\frac{1}{4}(\vartheta_1^2-\vartheta_2^2)(\vartheta_1^{}-\vartheta_2^{}) \,]^2_{}  \;. \\[3mm]
\end{array}
\end{equation}
Thus we have in the time domain
\begin{equation}
\langle\sigma_z^{}\tau_z^{}\rangle_t^{} \;=\; \sum_{i=1}^6 B_i^{}\,\rme^{\lambda_i^{}t}_{} \,+\, 
\langle\sigma_z^{}\tau_z^{}\rangle_{\rm eq}^{} \; ,
\end{equation}
where the $\{\lambda_j^{}\}$ are the zeros of $D(\lambda)=0$, and the equilibrium value is
\begin{equation}\label{sztzeq}
\langle\sigma_z^{}\tau_z^{}\rangle_{\rm eq}^{} \; =\;  - \frac{{\cal C}^{(-)}_{}(0)}{{\cal C}^{(+)}_{}(0)} 
\;=\; \frac{v}{2T}  \; .  
\end{equation}

The expressions (\ref{expecsigma}) -- (\ref{expectau}) and (\ref{expecsigtau}) -- (\ref{sigtauz3}) are the main results of
this work. They represent the exact analytical solutions for $\langle\sigma_z^{}(\lambda)\rangle$,
$\langle\tau_z^{}(\lambda)\rangle$, and $\langle\sigma_z^{}\tau_z^{}(\lambda)\rangle$, in the white-noise limit
for general coupling $v$ and general effective reservoir couplings $\vartheta_1^{}$ and $\vartheta_2^{}$.
Except for use of the form (\ref{wncorr}), no other approximation has been made. We remark that for all other initial and
final states of the RDM we would find the same pole functions (\ref{sz2}) and (\ref{sigtauz1}). 
Only the numerator function would be different.

\section{Qualitative features}

The behaviors of the four dynamical poles of $\langle\sigma_z^{}\rangle_t^{}$ and the six dynamical poles of
$\langle\sigma_z^{}\tau_z^{}\rangle_t^{}$, and the respective amplitudes are quite multifarious. 
In this section, we sketch the characteristics for the symmetric system,
$\bar\Delta_1=\bar\Delta_2\equiv\bar\Delta$ and $\vartheta_1^{}=\vartheta_2^{}\equiv \vartheta$. 

\subsection{ $\langle\sigma_z^{}\rangle_t^{}${\rm :}}

\begin{figure}[tp]
\centering
\includegraphics[width=6.0 cm, height=4.1 cm]{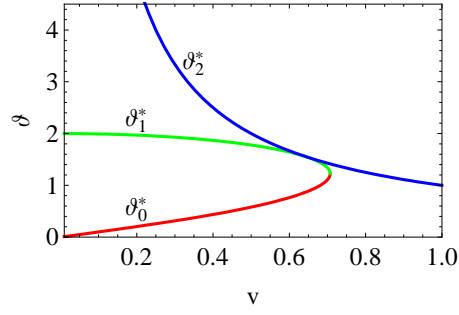}
\caption{$\langle\sigma_z^{}\rangle$: Crossover temperatures as functions of $v$, $\bar\Delta=1$.}
\label{fig:sz_crosstemp}
\end{figure}

\begin{figure}[tp]
\centering
\includegraphics[width=6.0 cm, height=4.1 cm]{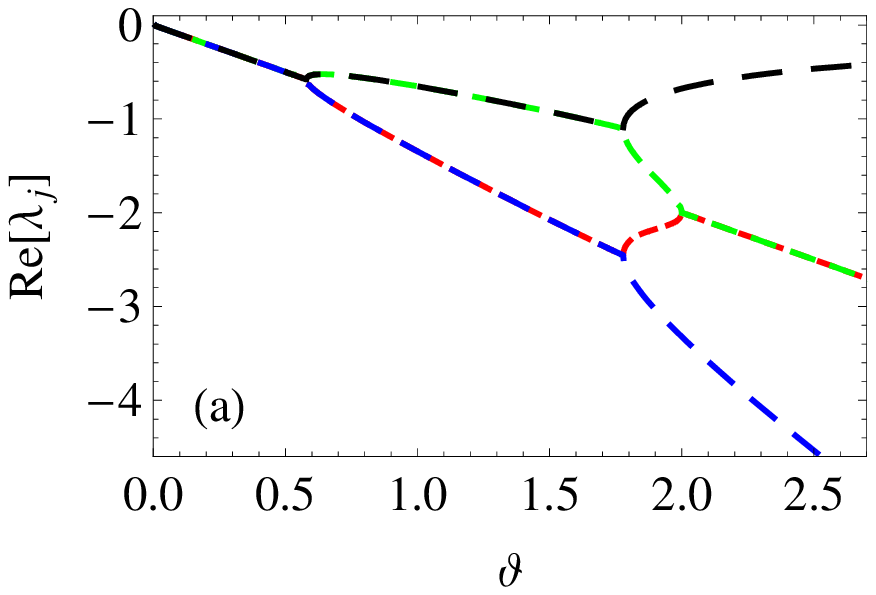}
\includegraphics[width=6.0 cm, height=4.1 cm]{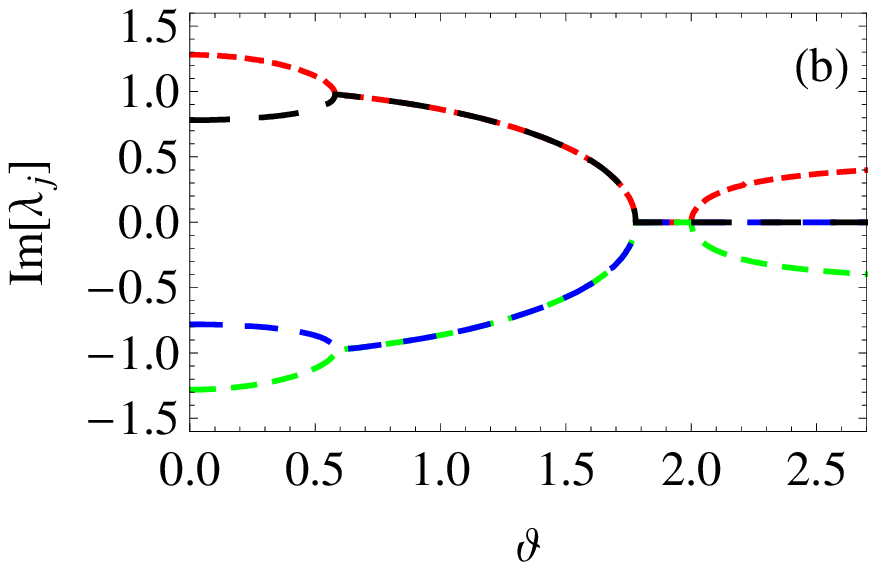}
\caption{$\langle\sigma_z^{}\rangle$: Real (a) and imaginary (b) part of $\lambda_j^{}$, $v = 0.5$, $\bar\Delta=1$.}
\label{fig:sz_re_im}
\end{figure}

In the coupling range $v< v_{\rm cr}^{} =\bar\Delta /\sqrt{2}$, there are three crossover temperatures, denoted by 
$\vartheta_0^\ast,\,\vartheta_1^\ast$, and $\vartheta_2^\ast$ (see Figs.~\ref{fig:sz_crosstemp}
and \ref{fig:sz_re_im}). 

In the  regime $\vartheta <\vartheta_1^\ast$ the dynamics is coherent and described by a superposition of two damped oscillations.
For $\vartheta <\vartheta_0^\ast$, the oscillations have different frequency and the same damping rate, and the amplitudes are
comparable in magnitude. On the other hand,
in the range $\vartheta_0^\ast < \vartheta < \vartheta_1^\ast$, they have the same frequency, but different
decrement and the amplitude belonging to the larger decrement is negligibly small.

In the temperature regime $\vartheta>\vartheta_1^\ast$, the dynamics is incoherent.
In the regime $\vartheta_1^\ast < \vartheta < \vartheta_2^\ast$, the four poles are real, 
and the two smallest rates have largest amplitudes and dominate the relaxation process.

In the so-called Kondo regime $\vartheta >\vartheta_2^\ast$, the dominant pole is real and approaches
$- \bar\Delta^2_{}/\vartheta$, and its residuum goes to $1 -\langle\sigma_z^{}\rangle_{\rm eq}^{}$, as temperature is increased. 
The other real pole takes the value $-2\vartheta$, while its residuum drops to zero.
There is also a damped oscillation of which the frequency  and rate approach asymptotically $v$ and $\vartheta$, but the amplitude 
becomes negligibly small. The phenomenon that, in the Kondo regime
for $K<\frac{1}{2}$, incoherent relaxation slows down with increasing temperature, is already well-known in the single 
spin-boson problem \cite{bookweiss}. 

\begin{figure}[tp]
\centering
\includegraphics[width=6.0 cm, height=4.1 cm]{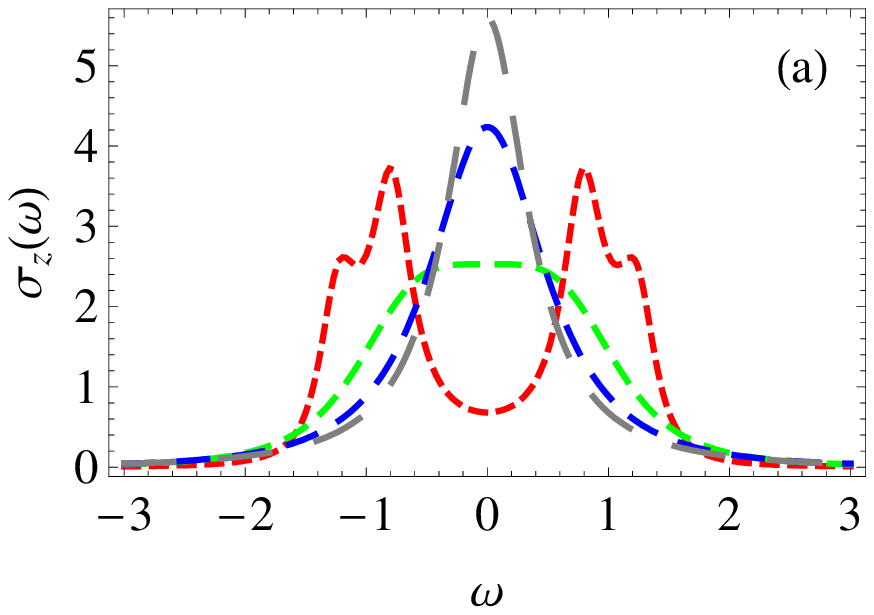}
\includegraphics[width=6.0 cm, height=4.1 cm]{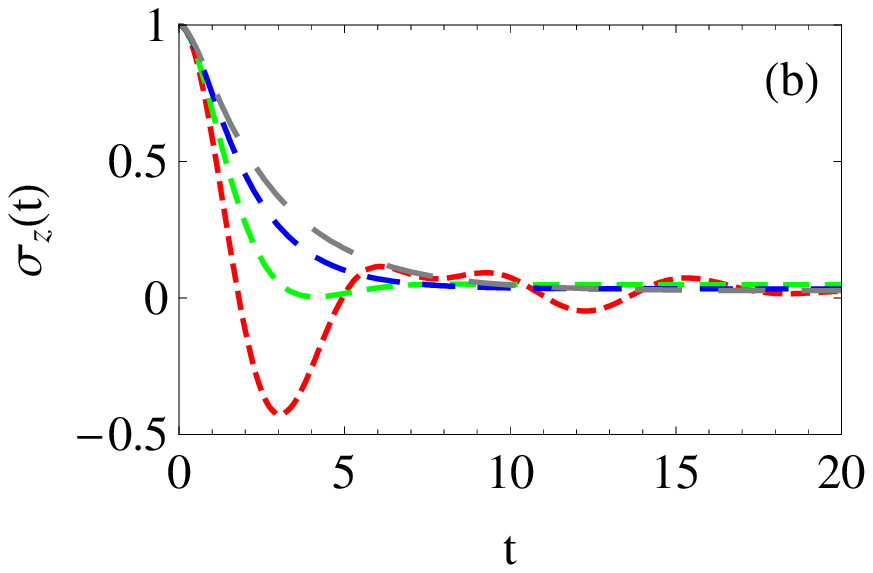}
\caption{$\langle\sigma_z^{}\rangle$: Plots of $\langle\sigma_z^{}\rangle$ in the Fourier regime (a) and time regime (b), 
$v = 0.5$, $\bar\Delta=1$, $K=0.05$.  
Red (small-dashed) $\vartheta=0.2$, green (dashed) $\vartheta=1.2$, blue (medium-dashed) $\vartheta=2.1$, 
grey (long-dashed) $\vartheta=2.8$.}
\label{fig:sz_fourier_time}
\end{figure}

Fig.~\ref{fig:sz_fourier_time} shows the transition from coherent to incoherent dynamics as $\vartheta$ is raised. One can also see
that at high $\vartheta$ the effective damping decreases with increasing $\vartheta$.

For $v>v_{\rm cr}^{}$, there is only one crossover. It separates the regime with two complex conjugate poles from the regime with one pair of complex conjugate poles and two real poles. Above $\vartheta^\ast_{}$, the relaxation is governed by the Kondo pole.

\subsection{ $\langle\sigma_z^{}\tau_z^{}\rangle_t^{}${\rm :}}

\begin{figure}[tp]
\centering
\includegraphics[width=6.0 cm, height=4.1 cm]{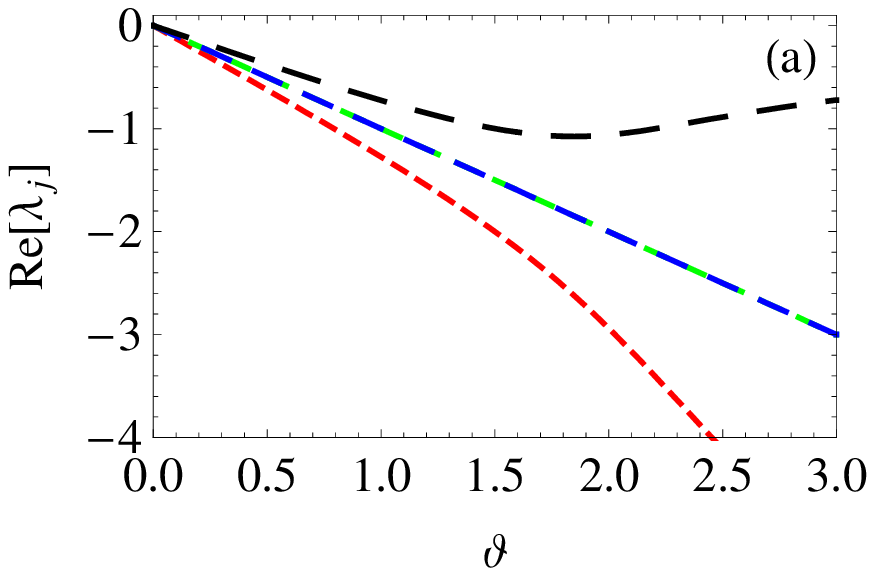}
\includegraphics[width=6.0 cm, height=4.1 cm]{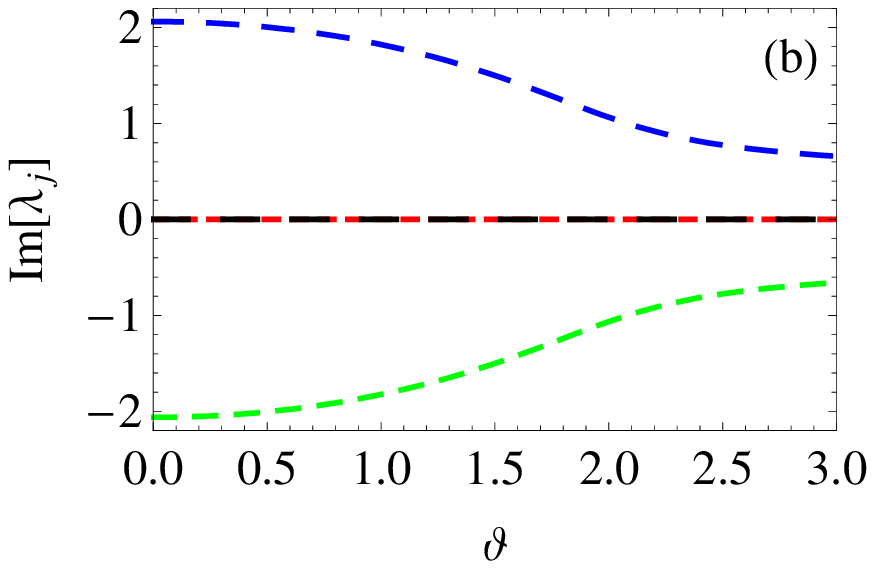}
\caption{$\langle\sigma_z^{}\tau_z^{}\rangle$: Real and imaginary part of $\lambda_j^{}$, $v = 0.5$, $\bar\Delta=1$.}
\label{fig:sztz_re_im}
\end{figure}

For a symmetric system, two poles are cancelled. The reduced pole equation reads
\begin{equation}
\fl\quad 
\lambda^4_{}\,+\, 4\vartheta\lambda^3_{}\,+\,(\bar\Omega_+^2+5\vartheta_{}^2)\lambda^2_{}\,+\,2\vartheta(\bar\Omega_+^2+\vartheta^2_{})\lambda
\,+\,4\vartheta^2_{}\bar\Delta^2_{}\;=\;0  \; .
\end{equation}
The characteristic behavior of the poles as function of the scaled temperature $\vartheta$ is shown in Fig.~\ref{fig:sztz_re_im}.
At low $\vartheta$, all poles contribute to the dynamics.
In the Kondo regime $\vartheta \gapx 3 \bar\Delta$, the leading pole  behaves as $- 2\bar\Delta^2_{}/\vartheta$, and the amplitudes
of the other contributions are negligibly small.

\section{Dynamics in the various parameter regimes for differing spins}

\subsection{Low temperature behavior}\label{subseclowtemp}

Below the first crossover temperature, $\Omega_\pm^{}\lapx T \le T_0^\ast$, the real parts of all poles are varying linearly with $T$. In detail, the poles and amplitudes are as follows.
\subsubsection{$\langle\sigma_z^{}\rangle_t${\rm :} }
There is a superposition of two damped oscillations with frequencies
\begin{equation}
\fl \quad
\begin{array}{rcl}
\lambda_{1,2}^{} &=& {\displaystyle \pm\,\mathrm{i}\,\bar\delta\,-\, \frac{\bar\Omega^2_{}+\bar\Delta_1^2- 2\,\bar\delta^2_{} }{2\,(
\bar\Omega^2_{} - \bar\delta^2_{})}\,\vartheta_1^{} \,-\, 
\frac{\bar\Omega^2_{}-\bar\Delta_2^2}{2\,(\bar\Omega^2_{} - \bar\delta^2_{})}\,\vartheta_2^{}\,+\,{\cal O}( \vartheta^2_{}) \; , }\\[3mm]
\lambda_{3,4}^{} &=&  {\displaystyle \pm\,\mathrm{i}\,\bar\Omega\,-\,\frac{2\,\bar\Omega^2_{} - \bar\delta^2_{} - 
\bar\Delta_1^2}{2\,(\bar\Omega^2 - \bar\delta^2)}\,\vartheta_1 \,-\, 
\frac{\bar\Delta_2^2-\bar\delta^2_{}}{2\,(\bar\Omega^2_{}- \bar\delta^2_{})}\,\vartheta_2^{} \,+\,{\cal O}( \vartheta^2_{}) \; , }
\end{array}
\end{equation}
and amplitudes 
\begin{equation}
\fl\quad
A_{1,2}^{} \;=\; \frac{1}{2}\,\frac{\bar\Delta_2^2 + v^2_{} -\bar\delta^2_{}}{\bar\Omega^2_{}-\bar\delta^2_{} 
} \,+\,{\cal O}( \vartheta)\; ,  \qquad
A_{3,4}^{} \;=\; \frac{1}{2}\,\frac{\bar\Omega_{}^2 - v^2_{} -\bar\Delta^2_{2}}{\bar\Omega^2_{}-\bar\delta^2_{}
 } \,+\,{\cal O}( \vartheta)\; .
\end{equation}
\subsubsection{$\langle\sigma_z^{}\tau_z^{}\rangle_t${\rm :} }
There is a superposition of two damped oscillations and two contributions describing incoherent relaxation
towards the equilibrium value $\langle\sigma_z^{}\tau_z^{}\rangle_{\rm eq}^{}$,
\begin{equation}
\fl\;
\begin{array}{rcl}
\lambda_{1,2}^{} &=& {\displaystyle \pm\,\mathrm{i}\,\bar\Omega_{+}^{} \,-\, \frac{\vartheta_1^{}\,+\,\vartheta_2^{}}{2}  \; , \qquad\quad\;\;\;
\lambda_{3,4}^{} \;=\;  \pm\,\mathrm{i}\,\bar\Omega_{-}^{}\,-\, \frac{\vartheta_1^{}\,+\,\vartheta_2^{}}{2}  \; , } \\[3mm]
\lambda_5^{} &=& {\displaystyle -\,\frac{\bar\Omega^2_{} - \bar\Delta _1^2}{\bar\Omega^2_{} - \bar\delta^2_{}} \vartheta_1^{} \,-\, 
\frac{\bar\Omega^2_{} -\bar\Delta_2^2}{\bar\Omega^2_{} - \bar\delta^2_{}} \vartheta_2^{}  \; , \quad
\lambda_6^{} \;=\; -\,\frac{\bar\Delta_1^2 - \bar\delta^2_{}}{\bar\Omega^2_{} - \bar\delta^2_{}} \vartheta_1^{} \,-\, 
\frac{\bar\Delta_2^2 - \bar\delta^2_{}}{\bar\Omega^2_{} - \bar\delta^2_{}} \vartheta_2^{}   \; ,  }
\end{array}
\end{equation}
where terms of order ${\cal O}(\vartheta^2_{})$ are disregarded. In leading order, the amplitudes are
\begin{eqnarray}
\fl\;
B_{1,2} &=& \frac{1}{4}\Big(\,1 - \frac{v^2_{}}{\bar\Omega_+^2}  \,\Big)\;, \qquad B_{3,4} \;=\; \frac{1}{4}\Big(\,1 - \frac{v^2_{}}{\bar\Omega_-^2}  \,\Big) \;,\\
\fl\;
B_5^{} &=& \frac{v^2_{}\bar\delta^2_{}}{(\bar\Omega^2_{}-\bar\delta^2_{})^2_{}}\, +\, 
\frac{\bar\delta^2_{}\,\langle\sigma_z^{}\tau_z^{}\rangle_{\rm eq}^{}}{
\bar\Omega^2_{}-\bar\delta^2_{}} \;,\qquad
B_6^{} \;=\; \frac{v^2_{}\bar\Omega^2_{}}{(\bar\Omega^2_{}-\bar\delta^2_{})^2_{}}\, -\, 
\frac{\bar\Omega^2_{}\, \langle\sigma_z^{}\tau_z^{}\rangle_{\rm eq}^{} }{ \bar\Omega^2_{}-\bar\delta^2_{}}\; .
\end{eqnarray}

 \subsection{The regimes of large coupling and/or high temperature} 
 When the coupling $v$ and/or the scaled temperatures $\vartheta_{1,2}^{}$ are large compared to the other frequencies, 
the amplitudes of three (five) pole contributions to $\langle\sigma_z^{}\rangle_t$ ($\langle\sigma_z^{}\tau_z^{}\rangle_t$) 
are negligibly small, and among the dynamical poles only the real pole with smallest
 modulus is relevant. Hence the two spins essentially  behave as a single spin which relaxes incoherently to the equilibrium 
state according to
\begin{equation}
\fl\qquad
\begin{array}{rcl}
\langle\sigma_z^{}\rangle_t &=& {\displaystyle \big(\, 1\,-\,\langle\sigma_z^{}\rangle_{\rm eq}\,\big)\,\rme^{-\gamma_\sigma^{}t}_{}
\, + \, \langle\sigma_z^{}\rangle_{\rm eq}^{}  } \; , \\[3mm]
\langle\sigma_z^{}\tau_z^{}\rangle_t &=& {\displaystyle \big(\, 1\,-\,\langle\sigma_z^{}\tau_z^{}\rangle_{\rm eq}\,\big)\,
\rme^{-\gamma_{\sigma\tau}^{}t}_{}
\, + \, \langle\sigma_z^{}\tau_z^{}\rangle_{\rm eq}^{}  } \; .
\end{array}
\end{equation} 

\subsubsection{$\langle\sigma_z^{}\rangle_t^{}${\rm :}} The relaxation rate is found from $D_1^{}(\lambda)=0$ with the 
form (\ref{sz2}) as
\begin{equation}\label{relaxrate}
\fl\qquad
\gamma_\sigma^{} \;=\; \frac{\bar\Delta_1^2}{v^2_{}+\vartheta_1^2}\,
\frac{\bar\Delta_2^2 +\vartheta_1^{}(\vartheta_1^{}+\vartheta_2^{})}{\vartheta_1^{}+\vartheta_2^{}} \; .
\end{equation}
This reduces in the parameter regime $\vartheta_1^{}\gg v,\,\bar\Delta_2^{}$ to
\begin{equation}\label{kondo1}
\fl\qquad
 \gamma_{\sigma}^{} \;=\; \bar\Delta_1^2/\vartheta_1^{} \; \propto  \; T^{2K_1 - 1}_{}  \; .
\end{equation}
In this regime, $\langle\sigma_z^{}\rangle_t^{}$ is independent of the coupling $v$ and hence independent of the 
dynamics of the $\tau$-spin.
The temperature dependence $\gamma_\sigma^{}\propto T^{2K_1^{}-1}_{}$ distinguishes the so-called Kondo regime, in which, for $K_1^{}<\frac{1}{2}$,
the relaxation dynamics slows down as temperature is increased.

On the other hand, when $v\gg \bar\Delta_{1,2}^{}$, the two spins are  locked together \cite{stamp}, and the effective tunneling matrix element is
$\bar\delta =\bar\Delta_1^{}\bar\Delta_2^{}/v$, as follows from (\ref{def_eigen1}) with (\ref{def_eigen2}). 
We then get from eq.~(\ref{relaxrate})
\begin{equation}
\fl\qquad
\gamma_\sigma^{} \;=\; \frac{\bar\delta^2_{}}{\vartheta_1^{}+\vartheta_2^{}}\,\Big[\,1 + \frac{\vartheta_1^{}
(\vartheta_1^{} +\vartheta_2^{})}{\bar\Delta_2^{2}}   \,\Big] \; .
\end{equation}
This yields the limiting expressions
\begin{equation}
\fl\quad 
\gamma_\sigma^{} \;=\; \frac{\bar\Delta_1^2}{v^2_{}}\,\vartheta_1^{} \quad (\bar\Delta_2^{}\ll \vartheta_{1,2}^{})\; ,
\quad\quad \;\mbox{and} \quad\quad \;  
\gamma_\sigma^{}\;=\; \frac{\bar\delta^2_{}}{ \vartheta_1^{}+\vartheta_2^{}} \quad
(\bar\Delta_2^{}\gg \vartheta_{1,2}^{})  \; .
\end{equation}
The former is the relaxation rate of the biased single spin-boson system at low $\vartheta_1^{}$. 
The latter describes Kondo-like joint relaxation of the locked spins.
\subsubsection{$\langle\sigma_z^{}\tau_z^{}\rangle_t^{}${\rm :} }
In the incoherent regime, the relaxation rate $\gamma_{\sigma\tau}^{}$ of the effective single spin 
receives rate contributions from both the $\sigma$- and the
$\tau$-spin as if these were
independent biased spins in contact with their own heat reservoir. 

In the large-coupling limit, $v \gg \bar\Delta_{1,2}^{},\,\vartheta_{1,2}^{}$, the relaxation rate is found as
\begin{equation}
\fl\qquad
\gamma_{\sigma\tau}^{} \;=\; \frac{\bar\Delta_1^2}{v^2_{}}\,\vartheta_1^{}\,+\, 
\frac{\bar\Delta_2^2}{ v^2_{}}\,\vartheta_2^{} \;.
\end{equation}
The individual contributions are single-spin rates in the large-bias regime.
In the high temperature limit $\vartheta_{1,2}^{}\gg v,\,\bar\Delta_{1,2}^{}$, on the other hand, both rate contributions are
Kondo-like,
\begin{equation}
\fl\qquad
\gamma_{\sigma\tau}^{} \; = \; \frac{\bar\Delta_1^2}{\vartheta_1^{}}\,+\,  \frac{\bar\Delta_2^2}{\vartheta_2^{}} \;.
\end{equation}

Consider next the regime $\vartheta_1^{}\gg \bar\Delta_1^{},\,\bar\Delta_2^{},\,v $, in which spin $\sigma$
behaves Kondo-like, as in eq.~(\ref{kondo1}). 
Hence the dynamics of the $\sigma$-spin is slow compared to that of the $\tau$-spin.
Thus we should expect that $\langle\sigma_z^{}\tau_z^{}\rangle_t$ approaches the dynamics of the 
biased single spin-boson case  as $\vartheta_1^{}$ is increased.
Taking into account terms of linear order in  $\gamma_\sigma^{}$ in the pole equation,
the expression (\ref{expecsigtau}) with (\ref{sigtauz2}) and 
(\ref{sigtauz1}) assumes the form
\[
 \fl 
 \langle\sigma_z^{}\tau_z^{}(\lambda)\rangle  =
\frac{\lambda^2_{}\,+\, 2\vartheta_2^{} \lambda  
\,+ \,v^2_{}+\vartheta^2_{2}  + \pi K_2^{} v\bar\Delta_2^2 /\lambda   }{ \lambda^3_{}\,+\, 2(\vartheta_2^{}+\gamma_\sigma^{}) \lambda^2_{}
\, + \, (\bar\Delta_{2}^2 + v^2_{}+\vartheta^2_{2}+ 3\gamma_\sigma^{}\vartheta_2^{})\lambda \,+\, \bar\Delta_2^2\vartheta_2^{}
+\gamma_\sigma^{}(v^2_{}+\vartheta_2^2) } \; .
\]
Indeed, in the limit $\gamma_\sigma^{}\to 0$, this form
reduces just to the analytic expression for $\langle\tau_z^{}(\lambda)\rangle$ of the biased single  spin-boson system in the 
white-noise limit \cite{bookweiss}. 

\section{Dynamics of a spin coupled to a spin-boson environment}
\begin{figure}[h!]
\centering
\includegraphics[width=6.0 cm, height=4.1 cm]{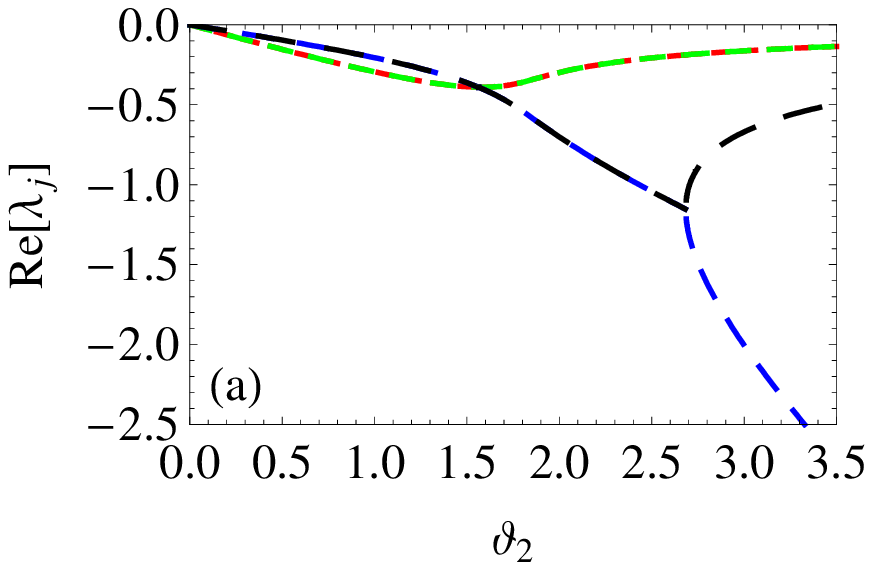}
\includegraphics[width=6.0 cm, height=4.1 cm]{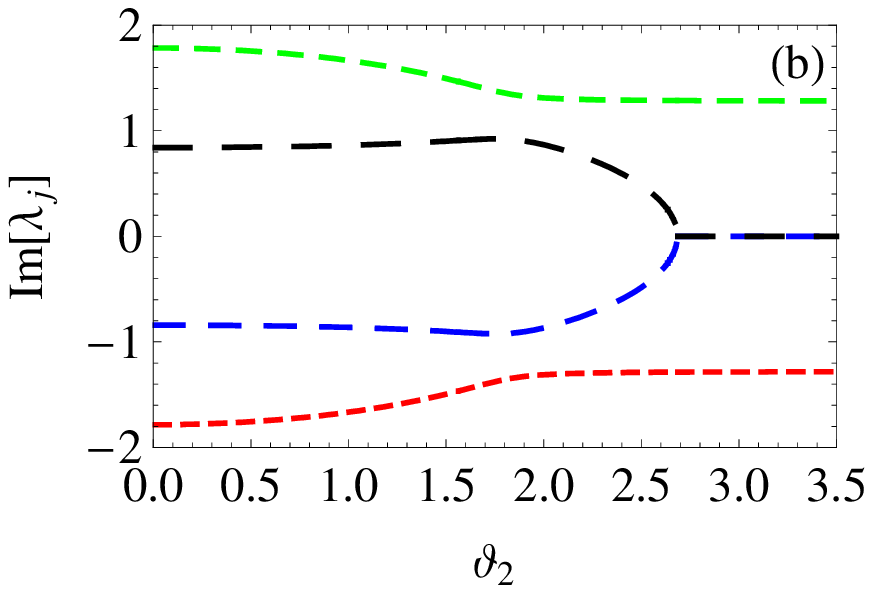}
\caption{$\langle\sigma_z^{}\rangle$ with spin-boson environment: Real (a) and imaginary (b) part of $\lambda_j^{}$. 
At low $\vartheta_2^{}$, $\langle\sigma_z^{}\rangle_t^{}$ is a superposition of 
two damped oscillations. At high $\vartheta_2^{}$, there is one damped oscillation and one relevant relaxation contribution.
The parameters are  $v = 0.8$, $\Delta_1=1$, $\bar\Delta_2=1.5$.}
\label{fig:sbe_re_im}
\end{figure}

Let us now view spin $\tau$ with reservoir 2 as an environment for spin $\sigma$.
This complex environment is in general non-Gaussian and non-Markovian \cite{paladino07}.
Recently, the same model has been studied numerically using a Markovian master equation approach \cite{bruder}.
To proceed, we first note that in the absence of bath 1,
$\vartheta_1^{}=0$, the pole equation $N_1^{}(\lambda)= 0$ is still 
of fourth order. There is no reduction in the general case. In Fig.~\ref{fig:sbe_re_im} we show plots of the four poles as
functions of $\vartheta_2^{}$ for a particular set of parameters.

\subsection{High temperature limit}
Simplification occurs, however, when $\vartheta_2^{}$ is very large compared to the other frequencies. 
In this regime, the kernel (\ref{kplus}) reduces to the form
 \begin{equation}\label{khtplus}
{\cal K}^{(+)}_{\rm ht}(\lambda) \;=\; - \lambda\,\frac{\Delta_1^2}{\lambda^2_{}+v^2_{}}
\left( 1\,+\, \frac{v^2_{}}{\lambda}\,\frac{\gamma_\tau^{}}{\lambda^2_{}+v^2_{}+ \gamma_\tau^{}\lambda}\right) \; ,
 \end{equation}
 where $\gamma_\tau^{} =\bar\Delta_2^2/\vartheta_2^{}$ is the relaxation rate of spin $\tau$ in the Kondo regime. 
With this high-temperature expression for the kernel, the quantity $\langle\sigma_z^{}(\lambda)\rangle$ is found to read
\begin{equation}\label{szsb}
 \fl\qquad
 \langle\sigma_z^{}(\lambda)\rangle \;=\; \frac{1}{\lambda \,+\, {\cal K}^{(+)}_{\rm ht}(\lambda)}\;=\;
 \frac{\lambda^2_{}+\gamma_\tau^{}\lambda+v^2_{}}{\lambda^3_{}+\gamma_\tau^{} \lambda^2_{} +
(\Delta_1^2+v^2_{})\lambda + \Delta_1^2\gamma_\tau^{}} \; .
 \end{equation}
This expression describes the dynamics of spin $\sigma$ coupled to a spin-boson environment, where the latter is in the Kondo regime.
To leading order in $\gamma_\tau^{}$, the poles of the expression (\ref{szsb}) are 
(see Fig.~\ref{fig:sbe_re_im} at large $\vartheta_2^{}$)
\begin{equation}\label{szsb1}
\fl\quad 
\lambda_{1,2}^{} \;=\;  \pm\,\rmi\,\sqrt{v^2_{}+\Delta_1^2}\;-\;\frac{v^2_{}}{2(v^2_{}+\Delta_1^2)}\,
\frac{\bar\Delta_2^2}{\vartheta_2^{}}   \; ,   \quad\quad
\lambda_3^{} \;=\;  -\,\frac{\Delta_1^2}{v^2_{}\,+\, \Delta_1^2}\,\frac{\bar\Delta_2^2}{\vartheta_2^{}}   \; ,  
\end{equation}
 and the amplitudes read
 \begin{equation}\label{szsb2}
 \fl\quad 
 A_{1,2}^{} \;=\;  \frac{\Delta_1^2}{2( v^2_{} + \Delta_1^2)} \,+\, {\cal O}\Big(\frac{1}{\vartheta_2^{} }\Big) \; ,  \quad\quad
A_{3}^{} \;=\; \frac{v^2_{}}{v^2_{} + \Delta_1^2} \,+\, {\cal O}\Big(\frac{1}{\vartheta_{2}^2}\Big) \; .  
\end{equation}
The expressions (\ref{szsb}) -- (\ref{szsb2}) may now be compared with the corresponding ones of a fictive single biased spin-boson system 
with parameters $\Delta_1^{}$ and $v$ in the white noise limit at scaled temperature $\tilde\vartheta$. The part that is symmetric 
in the bias reads
\begin{equation}\label{szsb3}
\fl \qquad
 \langle\sigma_z^{\rm (sb)}(\lambda)\rangle_{\rm s}^{} \;=\; 
  \frac{\lambda^2_{}\,+\, 2\tilde\vartheta\lambda\,+\, v^2_{}\,+\,\tilde\vartheta^2_{}}{\lambda^3_{}\,+\, 2\tilde\vartheta 
\lambda^2_{} \,+\,(\Delta_{1}^2+ v^2_{}+\tilde\vartheta^2_{})\lambda + \Delta_1^2\tilde\vartheta}  \; .
\end{equation} 
 We see that with the identification $\tilde\vartheta \entspricht \gamma_\tau^{} = \bar\Delta_2^2/\vartheta_2^{}$ the 
expressions (\ref{szsb}) and (\ref{szsb3}) are quite similar. 
Observe, however, that the damping rate of the oscillation is somewhat different because of the term 
$2\tilde\vartheta\lambda^2_{}$ in eq.~(\ref{szsb3}) instead of $\tilde\vartheta\lambda^2_{}$ in 
eq.~(\ref{szsb}) \cite{bookweiss}. Most importantly and interestingly in the correspondence, temperature  maps on 
the inverse of it.

\subsection{Linear response limit: spin-boson environment as a structured bosonic bath}
We should expect that, in the weak-coupling limit, the spin-boson environment is Gaussian and can be represented by a resonant power 
spectrum of a bath of bosons.
The Gaussian approximation of the spin-boson environment is found  by matching the power spectrum of the coupling of the $\sigma$-spin
 to the spin-boson environment \cite{paladino07} (with normalization as in eq.~(\ref{powerspec}))
\begin{equation}
\fl\qquad
 S_\tau^{}(\omega) \;=\; \frac{v^2_{}}{2\pi}\,{\rm Re}\,\int_{-\infty}^\infty \!\! \rmd t\,
\big\langle \,[\,\tau_z^{}(t)\tau_z^{}(0)\,+\, \tau_z^{}(0)\tau_z^{}(t)\, ]\,\big\rangle\, \rme^{{\rm i}\,\omega t}_{},  
\end{equation}
with that of a harmonic oscillator bath. In the white-noise limit of an unbiased spin, the symmetrized equilibrium correlation function
${\rm Re}\,\langle\tau_z^{}(t)\tau_z^{}(0)\rangle$ coincides with the expectation $\langle\tau_z^{}\rangle_t^{}$. Thus we obtain
\begin{equation}
\fl\quad
S_\tau^{}(\omega) \;=\; (2/\pi)\, v^2_{} \,{\rm Re}\,
\langle\tau_z^{}(\lambda=-\,\rmi\,\omega )\rangle \quad\;\mbox{with}\quad\;
\langle\tau_z^{}(\lambda)\rangle \;=\; \frac{\lambda +\vartheta_2^{}}
{ \lambda (\lambda +\vartheta_2^{}) +\bar\Delta_2^2 } \; .
\end{equation}
The resulting power spectrum is that of a structured bath of bosons with  a resonance of width $\vartheta_2^{}$ at frequency
$\omega= \bar\Delta_2^{}$,
\begin{equation}
S_\tau^{}(\omega) \;=\; \frac{2\vartheta_2^{}}{\pi}\,
\frac{ v^2_{}\bar\Delta_2^2}{(\bar\Delta_2^2 -\omega^2_{})^2_{}+ \vartheta_2^{2}\omega^2_{} } \; .
\end{equation}
Due to the coupling to the spin-boson environment, the spin $\sigma$ performs damped oscillation, $\langle\sigma_z^{}\rangle_t^{} =
\cos(\Delta_1^{}t)\,\rme^{-\gamma_{\rm dec}^{}t}_{}$. Upon calculating the decoherence rate in order $v^2_{}$, i.e.,
the so-called one-boson-exchange contribution of the effective boson bath, we obtain
\begin{equation}
\gamma_{\rm dec}^{} \;=\;\frac{\pi}{4}\, S_\tau^{}(\Delta_1^{}) \;= \; 
\frac{v^2_{}\bar\Delta_2^2 \,\vartheta_2^{}}{2\,[\,(\bar\Delta_2^2 -\Delta_1^2)^2_{}+ \vartheta_2^{2}\Delta_1^2 \,]} \; .
\end{equation}
 The analysis is completed by observing that this form emerges also upon calculating directly $\gamma_{\rm dec}^{}$
from the pole equation $D_1^{}(\lambda)=0$ with the form (\ref{sz2}).

\section{Conclusions}

We have studied the dynamics of a spin or qubit $\sigma$ coupled to another spin, which could be, for instance, another qubit, 
or a bistable impurity,
or a measuring device. We have solved the dynamics exactly  for white-noise reservoir couplings, and we have studied the rich behaviors
of the dynamics in diverse limits ranging from weak coupling and/or low temperatures to strong coupling and/or high temperature. 
We have also analyzed the effects of a spin-boson environment on the spin dynamics in the Gaussian and non-Gaussian domains. 

This paper has not attempted to
perform applications to already available experiments, instead we have tried to make some general points  
on complementary regimes and on the crossovers in between.   

One possible simple generalization beyond the white noise limit would be to replace the white-noise bath correlations
in time intervals in which the Laplace variable is irrelevant by the full quantum noise correlation, for instance 
\begin{equation}
\bar\Delta_\zeta^2\frac{\vartheta_\zeta^{}}{v^2_{}+\vartheta_\zeta^2}\quad\to\quad \Delta_\zeta^2 \cos(\pi K_\zeta^{}) 
\int_0^\infty\rmd s\,\cos(v s)\,\rme^{-Q'_\zeta(s)}_{} \;.
\end{equation}
The advantage would be two-fold: (i) the noise integral is known in analytic form for the Ohmic correlation function (\ref{ohmcorr}) \cite{bookweiss}, 
and (ii) with this substitution the algebraic form of pole equation and residua would be left unchanged.

One further generalization of the Hamiltonian (\ref{ham1}) is an applied bias acting on one or both of the spins, e.g.,
of the form $\epsilon_1^{}\sigma_z^{}$ and $\epsilon_2^{}\tau_z^{}$.  This important extra ingredient could be taken into account exactly in the
white noise regime, and would lead to additional shifts of the Laplace variable in all blip states of the $\sigma$-
and $\tau$-spin. Then one would end up with expressions of the form (\ref{expecsigma}) and (\ref{expecsigtau}) with
polynomials ramped up by bias terms. This extension will be discussed elsewhere. 

Finally, extension of the analysis of the dynamics to the regime $T\lapx\bar\Omega_\pm^{}$ requires to revert 
to the original expression (\ref{powerspec}) and compute its effect perturbatively in the one-boson-exchange approximation. 
This can be done either with the self-energy method presented in Ref.~\cite{bookweiss} or with the Redfield approach. 
One then find, e.g., that the actual equilibrium state of
$\langle\sigma_z^{}\tau_z^{}\rangle_{\rm eq}^{}$ for $K_\zeta^{} \ll 1$ is
\begin{equation}
\langle\sigma_z^{}\tau_z^{}\rangle_{\rm eq}^{} \;=\; \frac{v}{\bar\Omega^2_{} -\bar\delta^2_{}}\,\big[\,  
\bar\Omega\tanh(\beta\bar\Omega/2)\;-\; \bar\delta\tanh(\beta\bar\delta/2)  \,\big] \; . 
\end{equation}
This reduces for $T>\bar\Omega_\pm^{}$ to the previous form (\ref{sztzeq}) found in the white-noise regime.
The corresponding extension of the analysis and of the results given in Subsection \ref{subseclowtemp} to the domain
$T < \bar\Omega_\pm^{}$ will be reported elsewhere.

\section*{Acknowledgements} 
The authors wish to thank G. Falci and E. Paladino for valuable discussions. 
Financial support by the DFG through SFB/TR 21 is gratefully acknowledged.

\end{document}